\documentclass[sn-mathphys-ay, iicol]{sn-jnl}

\usepackage{graphicx}%
\usepackage{multirow}%
\usepackage{amsmath,amssymb,amsfonts}%
\usepackage{amsthm}%
\usepackage{mathrsfs}%
\usepackage[title]{appendix}%
\usepackage{xcolor}%
\usepackage{textcomp}%
\usepackage{manyfoot}%
\usepackage{booktabs}%
\usepackage{algorithm}%
\usepackage{algorithmicx}%
\usepackage{algpseudocode}%
\usepackage{listings}%

\theoremstyle{thmstyleone}%
\newtheorem{theorem}{Theorem}
\theoremstyle{thmstyletwo}%
\newtheorem{remark}{Remark}%

\theoremstyle{thmstylethree}%

\usepackage{times}
\usepackage{booktabs}
\usepackage{tablefootnote}

\usepackage{float}

\usepackage[binary-units=true]{siunitx}

\usepackage{amsmath}
\usepackage{amssymb}
\usepackage{amsthm}

\usepackage{graphicx}

\usepackage{amssymb}
\usepackage{pifont}

\usepackage{makecell}

\usepackage{wrapfig}

\usepackage{algorithm}
\usepackage{algpseudocode}

\usepackage{booktabs}

\usepackage{makecell}

\usepackage{pgfplots}
\usepackage{tikz}

\usepackage{float}

\usepackage{enumitem}

\usepgfplotslibrary{external}
\usetikzlibrary{shapes}
\usetikzlibrary{positioning}
\usetikzlibrary{arrows,decorations.markings}
\usetikzlibrary{calc,spy,shapes}
\usetikzlibrary{shapes}
\usepgfplotslibrary{fillbetween}

\usetikzlibrary{pgfplots.groupplots}
\usetikzlibrary{patterns}

\usepackage{caption}

\usepackage{subcaption}

\usepackage{xcolor}
\definecolor{applegreen}{rgb}{0.55, 0.9, 0.0}
\definecolor{amaranth}{rgb}{0.9, 0.17, 0.31}

\newtheorem*{theorem-non}{Theorem}
\newtheorem{lemma}{Lemma}
\newtheorem{corollary}{Corollary}
\newtheorem{assumption}{Assumption}

\newcommand{\fillopacity}{0.4}
\newcommand{\colordarkgreen}{green!50!black}

\newcommand{\lwboxplot}{0.7pt}

\newcommand{\myswarm}{\textsc{DMPC-Swarm}}

\newcommand{\opindent}{\phantom{=}}

\newcommand{\reviewerone}[1]{#1}

\newcommand{\reviewerthree}[1]{#1}
\newcommand{\reviewerall}[1]{#1}

\usepackage{ifthen}
	\newboolean{authnotes}
	\setboolean{authnotes}{true}

\ifthenelse{\boolean{authnotes}}
{
\newcommand{\thought}[1]{{\color[rgb]{0.2,0.39,0.66}(#1)}}
\newcommand{\todo}[1]{{\color[rgb]{1.0,0.0,0.0}(#1)}}
\newcommand{\fm}[1]{\footnote{{\bf\color{blue} Fabian: #1}}}
\newcommand{\mz}[1]{\footnote{{\bf\color{orange!50!black} Marco: #1}}}
\newcommand{\st}[1]{\footnote{{\bf\color{green!50!black} Sebastian: #1}}}
\newcommand{\ag}[1]{\footnote{{\bf\color{red!50!black} Alexander: #1}}}
}
{
\newcommand{\thought}[1]{}
\newcommand{\todo}[1]{}
\newcommand{\fm}[1]{}
\newcommand{\mz}[1]{}
\newcommand{\st}[1]{}
\newcommand{\ag}[1]{}
}

\newcommand{\capt}[1]{\mdseries{\emph{#1}}}

\newcommand{\ul}[1]{_\mathrm{#1}}
\newcommand{\uli}[1]{_{#1}}

\newcommand{\fakepar}[1]{\vspace{2mm}\noindent\textbf{#1}}

\pgfplotsset{
	box plot/.style={
		/pgfplots/.cd,
		black,
		only marks,
		mark=-,
		mark size=\pgfkeysvalueof{/pgfplots/box plot width},
		/pgfplots/error bars/y dir=plus,
		/pgfplots/error bars/y explicit,
		/pgfplots/table/x index=\pgfkeysvalueof{/pgfplots/box plot x index},
	},
	box plot box/.style={
		/pgfplots/error bars/draw error bar/.code 2 args={%
			\draw  ##1 -- ++(\pgfkeysvalueof{/pgfplots/box plot width},0pt) |- ##2 -- ++(-\pgfkeysvalueof{/pgfplots/box plot width},0pt) |- ##1 -- cycle;
		},
		/pgfplots/table/.cd,
		y index=\pgfkeysvalueof{/pgfplots/box plot box top index},
		y error expr={
			\thisrowno{\pgfkeysvalueof{/pgfplots/box plot box bottom index}}
			- \thisrowno{\pgfkeysvalueof{/pgfplots/box plot box top index}}
		},
		/pgfplots/box plot
	},
	box plot top whisker/.style={
		/pgfplots/error bars/draw error bar/.code 2 args={%
			\pgfkeysgetvalue{/pgfplots/error bars/error mark}%
			{\pgfplotserrorbarsmark}%
			\pgfkeysgetvalue{/pgfplots/error bars/error mark options}%
			{\pgfplotserrorbarsmarkopts}%
			\path ##1 -- ##2;
		},
		/pgfplots/table/.cd,
		y index=\pgfkeysvalueof{/pgfplots/box plot whisker top index},
		y error expr={
			\thisrowno{\pgfkeysvalueof{/pgfplots/box plot box top index}}
			- \thisrowno{\pgfkeysvalueof{/pgfplots/box plot whisker top index}}
		},
		/pgfplots/box plot
	},
	box plot bottom whisker/.style={
		/pgfplots/error bars/draw error bar/.code 2 args={%
			\pgfkeysgetvalue{/pgfplots/error bars/error mark}%
			{\pgfplotserrorbarsmark}%
			\pgfkeysgetvalue{/pgfplots/error bars/error mark options}%
			{\pgfplotserrorbarsmarkopts}%
			\path ##1 -- ##2;
		},
		/pgfplots/table/.cd,
		y index=\pgfkeysvalueof{/pgfplots/box plot whisker bottom index},
		y error expr={
			\thisrowno{\pgfkeysvalueof{/pgfplots/box plot box bottom index}}
			- \thisrowno{\pgfkeysvalueof{/pgfplots/box plot whisker bottom index}}
		},
		/pgfplots/box plot
	},
	box plot median/.style={
		/pgfplots/box plot,
		/pgfplots/table/y index=\pgfkeysvalueof{/pgfplots/box plot median index}
	},
	box plot width/.initial=1em,
	box plot x index/.initial=0,
	box plot median index/.initial=1,
	box plot box top index/.initial=2,
	box plot box bottom index/.initial=3,
	box plot whisker top index/.initial=4,
	box plot whisker bottom index/.initial=5,
}

\newcommand{\boxplot}[2][]{
	\addplot [box plot median,#1] table {#2};
	\addplot [forget plot, box plot box,#1] table {#2};
	\addplot [forget plot, box plot top whisker,#1] table {#2};
	\addplot [forget plot, box plot bottom whisker,#1] table {#2};
}

\usepackage{fancyhdr}
\begin{document} 

\title[DMPC-Swarm: Distributed Model Predictive Control on Nano UAV swarms]{DMPC-Swarm: Distributed Model Predictive Control on Nano UAV Swarms}

\author*[1]{\fnm{Alexander} \sur{Gräfe}}\email{alexander.graefe@dsme.rwth-aachen.de}

\author[1]{\fnm{Joram} \sur{Eickhoff}}\email{joram.eickhoff@dsme.rwth-aachen.de}

\author[2]{\fnm{Marco} \sur{Zimmerling}}\email{marco.zimmerling@tu-darmstadt.de}

\author[1]{\fnm{Sebastian} \sur{Trimpe}}\email{trimpe@dsme.rwth-aachen.de}

\affil[1]{\orgdiv{Institute for Data Science in Mechanical Engineering}, \orgname{RWTH Aachen University}, \orgaddress{\street{Theaterstraße 35-39}, \city{Aachen}, \postcode{52062}, \country{Germany}}}

\affil[2]{\orgdiv{Networked Embedded Systems Lab}, \orgname{TU Darmstadt}, \orgaddress{\street{Mornewegstraße 30}, \city{Darmstadt}, \postcode{64293}, \country{Germany}}}



\abstract{
	Swarms of unmanned aerial vehicles (UAVs) are increasingly becoming vital to our society, undertaking tasks such as search and rescue, surveillance and delivery. 
	A special variant of Distributed Model Predictive Control (DMPC) has emerged as a promising approach for the safe management of these swarms by combining the scalability of distributed computation with dynamic swarm motion control.
	In this DMPC method, multiple agents solve local optimization problems with coupled anti-collision constraints, periodically exchanging their solutions.
	Despite its potential, existing methodologies using this DMPC variant have yet to be deployed on distributed hardware that fully utilize true distributed computation and wireless communication.
	This is primarily due to the lack of a communication system tailored to meet the unique requirements of mobile swarms and an architecture that supports distributed computation while adhering to the payload constraints of UAVs.
	We present \myswarm{}, a new swarm control methodology that integrates an efficient, stateless low-power wireless communication protocol with a novel DMPC algorithm that provably avoids UAV collisions even under message loss. 
	By utilizing event-triggered and distributed off-board computing, \myswarm{} supports nano UAVs, allowing them to benefit from additional computational resources while retaining scalability and fault tolerance. 
	In a detailed theoretical analysis, we prove that \myswarm{} guarantees collision avoidance under realistic conditions, including communication delays and message loss.
	Finally, we present \myswarm{}'s implementation on a swarm of up to 16 nano-quadcopters, demonstrating the first realization of these DMPC variants with computation distributed on multiple physical devices interconnected by a real wireless mesh networks.
	A video showcasing \myswarm{} is available at \url{http://tiny.cc/DMPCSwarm}.

}

\maketitle


\clearpage
\newcommand{\distributed}[1]{\textcolor{magenta!50!black}{#1}}
\newcommand{\groundbased}[1]{\textcolor{green!50!black}{#1}}
\section{Introduction}

Robot swarms have the potential to transform fields like smart farming, civil protection, and planetary exploration~\citep{trianni2016saga, blender2016xaverproject, cantizani2022bluetoothsearchandrescue, couceiro2013collectivesearchandrescue, staudinger2021robotswarmcommunication, zhang2020autonomousswarmnavigation, kang2019marsbee}. 
Although currently confined to laboratories, they are expected to become commonplace by 2030~\citep{dorigo2020reflections}.
This holds especially for swarms of unmanned aerial vehicles (UAVs) relevant for tasks such as smart farming~\citep{Walter2017, trianni2016saga, r2018research}, aerial surveillance~\citep{saska2016swarm, chung2018survey}, and wildlife observation~\citep{shah2020multidrone}. 

A fundamental principle in robot swarms is \emph{distributed control}, where agents collaboratively make decisions through distributed computation and communication, unlike \emph{centralized control} by a single agent. 
Distributed control leverages the collective's computational power, enhances scalability, and improves resilience by avoiding single points of failure~\citep{mondada2004swarm, Luis2019, ge2017distributed}.

Building upon these advantages, recent work proposed a distributed optimization approach based on \emph{Distributed Model Predictive Control} (DMPC), which is one of the most promising distributed swarm control methods in recent years~\citep{Luis2019,Luis2020,Park2022,Park2023,Chen2023,Chen2022,Graefe2022}.
These approaches enable dynamic swarm maneuvers by continuously solving distributed optimization problems across the agents. 
Moreover, by incorporating specific constraints into its optimization problem, they can formally guarantee collision avoidance~\citep{Park2022, Park2023, Chen2023, Chen2022, Graefe2022}, which is crucial for ensuring safety in swarm operations.


However, despite promising theoretical analyses and simulations demonstrating DMPC's potential, existing hardware implementations execute DMPC on a central computer, while simulating selected effects of distributed computation and communication~\citep{Luis2020,Park2022,Park2023,Chen2023,Chen2022,Graefe2022}. 
Although these implementations confirm DMPC's conceptual suitability for swarm control, they do not realize the practical benefits of distributed architectures. 
Moreover, centralized implementations oversimplify real-world conditions, particularly communication delays and message loss, rendering them inadequate for evaluating DMPC-controlled swarms in scenarios where the advantages of distributed systems are essential.

To bridge this gap, we develop and implement \myswarm{}, the first \emph{distributed realization} of DMPC for swarm control over wireless networks. 
We identify three key challenges to a real-world realization of DMPC, which we address in this work:

\begin{enumerate}
    \item[\textbf{C1}] \textbf{Communication:} 
    DMPC requires communication between agents over a wireless network. 
    In \myswarm{}, we aim to achieve this communication over self-organizing wireless mesh networks, where agents transmit data for one another based on device-to-device communication. 
    This preserves the distributed nature of the swarm and enhances scalability, reliability, and efficiency compared to a network in which agents are limited to direct communicate with a dedicated base station~\citep{laneman04}.
    However, wireless mesh networks introduce complexities, especially as robots move and the network topology continuously evolves~\citep{khan2019hybrid, oubbati2019routing, namdev2021optimized}. 
    Consequently, achieving the communication required by DMPC approaches becomes even more challenging.

    \item[\textbf{C2}] \textbf{Computation:} 
    DMPC necessitates significant computational power because it involves repeatedly solving optimization problems. However, UAVs, especially nano-quadcopters used in applications such as agriculture~\citep{gago2020nano}, indoor source seeking~\citep{duisterhof2021tiny, karaguzel2023shadows} and indoor visual inspection~\citep{tavasoli2023real}, have limited onboard computing capabilities due to constraints in size, weight, and cost.
    Despite these limitations, most existing DMPC implementations fail to address the computational restrictions inherent in such UAVs. Furthermore, swarm systems often comprise heterogeneous computing architectures with varying computational capabilities, for instance, quadcopters of different sizes, mobile robots, and operator devices. Current DMPC approaches generally do not exploit this diversity in computational resources.
    
    \item[\textbf{C3}] \textbf{Unrealistic assumptions of DMPC:} 
    Moreover, existing DMPC methods often overlook practical challenges such as computational delays due to limited processing power, communication delays and message loss resulting from unreliable wireless communication. 
    These issues pose significant hurdles, as they may cause the collision avoidance guarantees provided by DMPC methods to fail in practice, thereby limiting their real-world applicability. 
    Solving these challenges on the algorithmic side is crucial for the successful deployment of DMPC in practical swarm applications.

\end{enumerate}

We address these challenges through a novel combination of a stateless communication protocol, a distributed and event-triggered compute architecture leveraging multiple heterogeneous compute nodes, as well as algorithmic extensions to DMPC, achieving the first distributed realization of DMPC.

\reviewerthree{We note that different concepts and notions of distributed systems exist. In our work, we define ``distributed'' as collaborative problem-solving among multiple agents, sharing workloads and operating without a single coordinator~\citep{hu2018centralize,cheraghi2022pastpresentfuture,lupashin2014flyingmachinearena}. 

One prevalent alternative view on distributed systems emphasizes interaction and information exchange with agents in a neighborhood and peer-to-peer fashion~\citep{antonelli2013interconnected, ge2017distributed}. 
Although such approaches offer scalability benefits, they can also involve downsides in other settings.
For example, defining and tracking which agents qualify as neighbors is often problematic, particularly in scenarios involving rapidly moving swarms, which complicates the theoretical analysis of algorithms.
This is especially relevant when the interaction involves wireless communication, where the behavior and existence of communication links to neighbors can be difficult to predict due to the significant influence of environmental factors. 

Following our view on ``distributed'' as collaborative problem-solving and distribution of workloads, we consider any system that facilitates these functions as a potential solution, rather than restricting interaction to a peer-to-peer fashion.
\reviewerall{Accordingly, we design and integrate all three system components, algorithms, computing infrastructure, and communication infrastructure, into an architecture featuring distributed entities at each component.}
For example the distributed communication solution that we propose is based on physical neighbor-to-neighbor communication, while enabling a many-to-all information exchange and global interaction as part of the solution.

}



Before detailing our contributions, we make the swarm control problem precise.

\subsection{Problem Setting}
\label{sec:introduction:problemsetting}



We consider a swarm $\mathcal{A}=\{1,\cdots,N\}$ of $N$ UAVs.  
At any time $t$, each UAV $i \in \mathcal{A}$ has a position $p_{i}(t) \in \mathbb{R}^3$. 
A UAV can measure its own position but not the positions of the other UAVs.
Each must navigate from its initial position $p_{i, \mathrm{init}} = p_{i}(0)$ to a target position $p_{i, \mathrm{target}}$ without colliding with other UAVs.

The target positions $p_{i, \mathrm{target}}$ may change over time, e.g., when a UAV receives a new task at a different location. 
This dynamic environment necessitates frequent real-time trajectory replanning.
Such distributed position-following problem is the basis of many swarming tasks and is the standard problem setting of DMPC~\citep{Luis2019,Luis2020,Park2023,Park2022,Chen2023,Chen2022}.

Each UAV can run a low-level trajectory-following controller but lacks the resources to solve the optimization problems in DMPC. 
However, the network---in addition to the $N$ UAVs---also includes $M$ agents ($M > 1$) with higher computational power, called \emph{compute units} (CUs). 
We assume $M < N$ to account for limited capacities. 
Such CUs are common in practice, e.g., in smart farming or manufacturing, UAVs may connect to digital devices (e.g., smartphones or laptops) or rely on edge clouds and computational resources from other agents like ground robots or larger UAVs~\citep{baumann2020wireless, sankaranarayanan2023paced, liu2021boost}.

All UAVs and CUs communicate via a wireless mesh network, each device equipped with a wireless transceiver. 
Given the limited range of these transceivers, devices can directly communicate only with their immediate neighbors. 
As the agents move, the network topology changes dynamically, continuously altering the quality and availability of direct communication links, which is known to be challenging for efficient and reliable communication~\citep{khan2019hybrid, oubbati2019routing,namdev2021optimized,chriki2019fanet,lv2023survey}.

\textbf{Problem Statement.} Our goal is to develop a swarm architecture for the safe control of real UAV swarms using DMPC, realized on distributed physical hardware and a real wireless mesh network. 
The DMPC computations must be distributed across multiple CUs, considering their limited number, and should ensure that the UAVs' positions converge to the target positions over time
\begin{equation}
 \forall i\in \mathcal{A}: p_{i}(t)\to p_{i,\mathrm{target}},
\end{equation}
while provably avoiding collisions between UAVs
\begin{align} 
    \label{eq:truecollisionavoidance} 
    \forall t\in\mathcal{R}_0^+,&~\forall i,j\in\mathcal{A},~i\neq j:\\ 
    &\left\| \Theta^{-1}[p_j(t)-p_i(t)] \right\|_2 \geq d_\mathrm{min},\nonumber 
\end{align}
where $\Theta$ is a scaling matrix for downwash effects, and $d_\mathrm{min}$ is the minimum allowable distance~\citep{Luis2019, Luis2020, Chen2022, Chen2023,Park2022, Park2023, Graefe2022}.
The DMPC framework should function under real-world conditions, such as communication message loss, delays and limited computational resources.

\subsection{Contributions}

Our solution, \myswarm{}, effectively solves this problem by addressing challenges \textbf{C1}--\textbf{C3}.
Overall, we make the following contributions:

\begin{enumerate}
    \item We present \myswarm{}, a novel swarm architecture combining low-power wireless communication, distributed computing and a new DMPC algorithm. 
    The communication protocol is resilient to rapid device movement and provides the communication required by DMPC.
    The architecture building on top of it efficiently balances the DMPC computational load across UAVs and CUs. 

    \item \reviewerthree{We introduce message-loss-recovery DMPC (MLR-DMPC) as the algorithmic part of \myswarm{}.}. This DMPC method handles message loss and communication delays, providing collision avoidance guarantees under realistic conditions.

    \item We implement \myswarm{}, achieving the first distributed hardware realization of DMPC-based swarm control on nano-quadcopters. 
    Our experimental setup features up to 16 nano-quadcopters communicating over a wireless mesh network using Bluetooth Low Energy (BLE). 
    The entire system is based on the popular Crazyflie platform, with its software and hardware files accessible at \url{https://github.com/Data-Science-in-Mechanical-Engineering/DMPC-Swarm}.
    A video demonstrating our results is available at \url{http://tiny.cc/DMPCSwarm}.
\end{enumerate}
\section{Related Work}
\label{sec:relatedwork}

Before presenting our solution, we analyze related DMPC methods. 
For comprehensive reviews of general UAV swarm control methods, we refer to~\cite{gugan2023path, wang2024review, yu2023overview}.

\label{sec:introduction:background}

 \reviewerthree{
    DMPC encompasses a broad category of methods.
    \cite{peng2024distributed} provide an overview of various DMPC types specifically applied to swarm control.
    Among these variants, the approaches presented in \cite{Luis2019,Luis2020,Park2022,Park2023,Chen2023,Chen2022,Graefe2022} form the backbone of our work.
    For brevity, we refer to these approaches simply as DMPC throughout the paper.

    It is important to emphasize that alternative DMPC methods also exist. For instance, some approaches solve centralized optimization problems using distributed techniques~\citep{stomberg2021distributed,stomberg2023dmpc,camponogara2002distributed}. In contrast to the DMPC variant employed here, these methods typically confine information exchange to the local neighborhood of agents.

The considered DMPC variant originated from \cite{augugliaro2012generation}, which introduced a centralized Sequential Convex Programming approach for trajectory planning by linearizing nonlinear collision constraints and solving Quadratic Programs (QPs).}
\cite{Luis2019} extended this method to distributed computation by assigning one QP per UAV; after solving their QPs and simulating a step forward, the swarm executes the final trajectories together. 
\cite{Luis2020} adapted the method for real-time control, synchronizing UAVs into periodic rounds.
During such a round, each UAV first runs a QP.
Afterwards, the UAVs exchange their solution in many-to-all communication rounds.
In parallel, each UAV executes its initial trajectory segment. 

Subsequent works extended DMPC to provide theoretical collision avoidance guarantees.
\cite{Park2022} utilized Bernstein polynomials for trajectory representation, leveraging their convex hull properties and final state constraints to ensure collision avoidance, recursive feasibility, and static obstacle avoidance. 
\cite{Park2023} extended this approach to dynamic obstacles, mitigating deadlocks through heuristics that temporarily adjust target positions.
Similarly, \cite{Chen2022} showed that soft constraints can prevent deadlocks with theoretical guarantees, which \cite{Chen2023} extended to include static obstacles. Both use time-varying separating planes, the dynamic extension of Buffered Voronoi Cells (BVC)~\citep{Zhou2017}, and final state constraints for collision avoidance and recursive feasibility.
Concurrently, \cite{Graefe2022} studied the feasibility of DMPC for nano UAV swarms by proposing to offload computations to ground agents and using event-triggered scheduling, however, without a distributed hardware implementation.
Further, they employ time-varying BVCs with final state constraints for collision avoidance guarantees.

In addition to theoretical advancements, the presented works' hardware experiments have demonstrated DMPC's practical suitability.
However, these rely on a \emph{single} central computer that simulates distributed computation and communication, streaming position commands to the UAVs.
This setup fails to capture real-world conditions like communication delays from limited communication bandwidth, message loss, and constrained computational power.

In contrast, we present a distributed hardware implementation.
Building upon \cite{Graefe2022}, we employ ground-based distributed event-triggered computation. 
To ensure collision avoidance even in the presence of message loss, a common occurrence in wireless communication, we propose MLR-DMPC, an extension of DMPC. 
Furthermore, to effectively mitigate deadlocks, we incorporate soft constraints~\citep{Chen2022, Chen2023} and high-level planning heuristics~\citep{Park2022, Park2023}.

Finally, we distinguish these methods from others also termed DMPC~\citep{peng2024distributed}.
In \cite{stomberg2021distributed, stomberg2023dmpc}, for example, centralized optimization problems are solved using distributed techniques.
While effective for controlling ground robots via distributed off-board computations, these methods require multiple communication steps per optimization run, making it too slow when running under limited communication bandwidth.

\section{\myswarm{} --- Architecture}
\noindent
This section presents \myswarm{}, providing an overview of how it enables communication (\textbf{C1}), distributed computation (\textbf{C2}), and safe control (\textbf{C3}). 
The subsequent Section~\ref{sec:mlrdmpc} details the methodologies of \myswarm{}.

\begin{figure*}[h]
    \centering
    \includegraphics[width=0.9\textwidth]{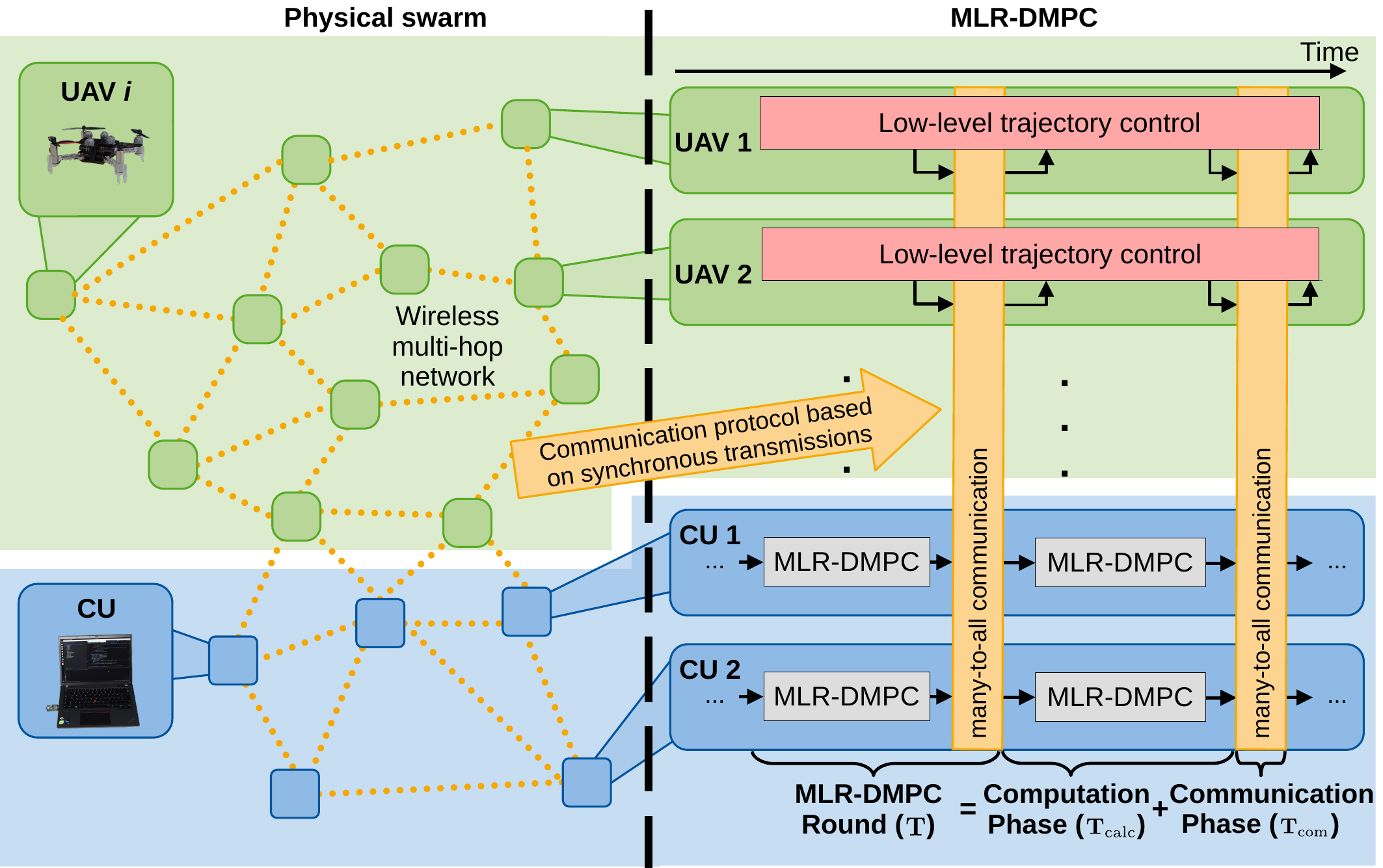}

	\caption{\myswarm{} overview.
    \capt{\textbf{Left:} The physical swarm consisting of of UAVs and CUs connected via a wireless mesh network. \textbf{Right:} Swarm operations are structured in synchronized rounds, alternating between computation and many-to-all communication phases, facilitated by Mixer.}}

    \label{fig:approach}
\end{figure*}

\label{sec:method:overview}

\subsection{Communication}
As described in Section~\ref{sec:relatedwork}, DMPC requires synchronized round-based many-to-all communication.
However, the inherent unreliability of wireless networks, combined with the high dynamics of moving UAVs, makes this challenging.


Over the past decade, wireless mesh protocols based on synchronous transmissions have proven superior to traditional point-to-point routing approaches~\citep{zimmerling20st}.
The key insight is that packet collisions from transmission that overlap in time, space and frequency, can be successfully decoded due to the capture effect~\citep{Leentvaar1976} and non-destructive interference~\citep{herrmann22rssispy}.
This has two crucial implications: (\emph{i}) Unlike traditional routing protocols, synchronous transmission methods do not need to track topology changes to avoid packet collisions, making them highly resilient to such changes. (\emph{ii}) Synchronous transmissions enable distributed network nodes to achieve efficient time synchronization with sub-microsecond accuracy~\citep{glossy,chaos}, providing the foundation for real-time communication with formally proven end-to-end deadline guarantees~\citep{zimmerling17realtime}. 
As a result, protocols based on synchronous transmissions have enabled a range of powerful wireless control applications~\citep{FeedbackControlGoesWireless, baumann2019fast, PredictiveTriggering, Trobinger2021}.


In \myswarm{}, we build on this prior work by leveraging a synchronous transmissions based protocol called Mixer~\citep{Mixer}. 
Mixer provides many-to-all communication across dynamic wireless mesh networks, essential for DMPC.
It does this also efficiently: Mixer achieves order-optimal scaling with the number of messages by integrating synchronous transmissions with random linear network coding~\citep{Ho2006}.
The network coding approach also provides additional reliability in real networks with fast-moving nodes like UAVs. 
This high robustness and reliability allows us to treat message losses as exceptional events when designing the control system in \myswarm{}.

In \myswarm{}, Mixer synchronizes the devices into discrete-time rounds indexed by~$k$ and with a fixed duration~$T$ (see Figure~\ref{fig:approach}).
Each round comprises a computation and a communication phase, which last~$T_\mathrm{calc}$ and~$T_\mathrm{com}$, respectively. 
Devices begin computation simultaneously at~$t = kT$ and begin to communicate at $t = kT + T_\mathrm{calc}$. During communication, devices concurrently send messages in dedicated, synchronized time slots.
Mixer efficiently broadcasts these messages, all devices can receive the complete set after $T_\mathrm{com}$. This mechanism provides the timely, reliable data exchange essential for MLR-DMPC.

\subsection{Distributed Computation}

Nano-UAV swarms face a fundamental conflict between limited onboard compute resources and the high computational demands of DMPC. 
To address this, we leverage the swarm's $M$ CUs for heavy computations (cf.\ Section~\ref{sec:introduction:problemsetting}). 

Although each CU can compute a trajectory for one UAV at once via DMPC, maintaining a one-to-one ratio of CUs to UAVs ($M = N$) is inefficient---it is costly, overloads the compute network, and demands high communication bandwidth. 
Therefore, \myswarm{} employs significantly fewer CUs ($M < N$). 
In each round~$k$, the CUs compute new trajectories for $M$ of the $N$ UAVs, while the rest follow their previously planned paths. 
A distributed, priority-based event trigger determines which UAVs need new trajectories and schedules computations on the CUs (see Section~\ref{sec:mlrdmpc:algorithm:eventtrigger}).

\subsection{Overview of MLR-DMPC}
\label{sec:method:mldrmpc}

\myswarm{} leverages Mixer's round-based structure of alternating computation and communication phases for MLR-DMPC (see Figure~\ref{fig:approach}). 
A key innovation of MLR-DMPC over existing DMPCs is its ability to handle message loss. 
We provide a brief overview here; details are in Section~\ref{sec:mlrdmpc}.

To ensure effective DMPC, the CUs must be informed of all UAV actions, which Mixer's many-to-all communication facilitates.
The CUs monitor received UAV activities using information trackers.

At the start of each computation phase, each CU checks whether its information trackers are up-to-date. 
If they are, the CU solves a QP for its assigned UAV and broadcasts the solution during the next communication phase. 
If crucial information is missing---indicating a critical message loss that prevents safe trajectory calculation---the CU does not proceed. 
Instead, the CU identifies the missing information and requests it in the subsequent communication phase.
Although such instances are rare due to Mixer's reliability, they must be addressed. 

Consequently, UAVs receive only recursively feasible trajectories that ensure their safety—even if they do not receive an update due to message loss or not being selected by the trigger. 
The UAVs track these trajectories using high-frequency, low-level controllers common in DMPC~\citep{Park2022, Park2023, Chen2023, Chen2022, Graefe2022}. 



\section{Distributed Model Predictive Control with Message Loss Recovery}
\label{sec:mlrdmpc}

This section details MLR-DMPC. 
After describing the UAV model forming the base for MLR-DMPC, we provide an in-depth explanation of MLR-DMPC and prove collision avoidance.

\subsection{UAV Model}

All $N$ UAVs have nonlinear dynamics
\begin{align}
    \label{eq:uavdynamics}
    \forall i \in \mathcal{A}:\quad&\dot{x}_i(t) = f_i(x_i(t), u_i(t)) + v_i(t),\nonumber\\ &x_i(0) = x_{i,0},
\end{align}
where $x_i \in \mathcal{X} \subseteq \mathbb{R}^n$ is the state vector, $u_i \in \mathcal{U} \subseteq \mathbb{R}^m$ is the control input, $f_i: \mathcal{X}\times\mathcal{U}\to\mathcal{X}$, and $v_i \in \mathcal{V} \subseteq \mathbb{R}^n$ represents disturbances.
The position of UAV $i$ at time $t$ is given by $p_i(t) = g_{\mathrm{p}, i}(x_i(t))$, where $g_{\mathrm{p}, i}: \mathcal{X} \rightarrow \mathcal{P} \subseteq \mathbb{R}^3$ maps the state to the position.
Typically, $x_i$ includes the position, making $g_{\mathrm{p}, i}$ straightforward (e.g., $p_i(t) = [I, 0] x_i(t)$).
For detailed quadcopter dynamics, see~\citep{mahony2012multirotor, antal2023modelling, panerati2021learning}.
Each UAV must navigate from its initial position $p_{i, \mathrm{init}} = g_{\mathrm{p}, i}(x_{i,0})$ to its target $p_{i, \mathrm{target}}$ while avoiding collisions (Equation~\ref{eq:truecollisionavoidance}).

\reviewerall{
Instead of using the complex UAV dynamics $f_i$, we define a simpler nominal system for each UAV $i \in \mathcal{A}$, as common in (D)MPC~\citep{augugliaro2012generation, Luis2019, Luis2020, Chen2022, Chen2023, Graefe2022}
\begin{equation}
    \label{eq:nominalsystem}
    \dot{\hat{x}}_i = \hat{f}_i(\hat{x}_i, \hat{u}_i),\quad \hat{x}_i(0) = \hat{x}_{i,0},
\end{equation}
where $\hat{x}_i \in \hat{\mathcal{X}}$ is the nominal state with $\hat{x}_{i,0}$ its initial value, $\hat{u}_i\in\hat{\mathcal{U}}$ and $\hat{f}_i:\hat{\mathcal{X}}\times\hat{\mathcal{U}}\to\hat{\mathcal{X}}$.
We define the function $\hat{g}_{\mathrm{p}, i}:\hat{\mathcal{X}}\to\mathcal{P}$ that maps the nominal state to the nominal position $\hat{p}_i(t) = \hat{g}_{\mathrm{p}, i}(\hat{x}_i(t))$.

We further design a controller $c_i: \mathcal{X} \times \hat{\mathcal{X}} \times \hat{\mathcal{U}} \rightarrow \mathcal{U}$ that controls the UAV
\begin{equation}
    \dot{x}_i = f_i\big(x_i,c_i(x_i, \hat{x}_i, \hat{u}_i)\big) + v_i,
\end{equation} 
steering it towards the nominal state.
In the following, we will not explicitly specify $\hat{f}_i$ and $c_i$, instead, we will make the following assumption similar to incremental asymptotic stability~\citep{kohler2018novel}.

\begin{assumption}
\label{as:tracking}

We assume that for arbitrary inputs $\hat{u}_i \in \hat{\mathcal{U}} \subseteq \mathbb{R}^{\hat{m}}$, there exists a constant $\Delta d_\mathrm{min}\in\mathbb{R}_{\geq 0}$ such that if $||\hat{p}_i(0)-p_i(0)||=||\hat{g}_{\mathrm{p}, i}(\hat{x}_i(0))-g_{\mathrm{p}, i}(x_i(0))||\leq \Delta d_\mathrm{min}$,
then for all $t\in\mathbb{R}_{\geq 0}$:
\begin{equation}
    \left\| p_i(t) - \hat{p}_i(t) \right\|_2 \leq \Delta d_\mathrm{min}.
\end{equation}
\end{assumption}
}


We can satisfy this assumption by selecting $\hat{f}_i$ as a fourth-order integrator in combination with an appropriate controller leveraging the differential flatness of quadcopters~\citep{mellinger2011minimum}. 
However, empirical studies show that second- or third-order integrators suffice for generating smooth, trackable trajectories while reducing QP complexity~\citep{augugliaro2012generation, Luis2019, Luis2020, Chen2022, Chen2023, Graefe2022}. 
Therefore, \myswarm{} uses a third-order integrator for $\hat{f}_i$, with input $\hat{u}_i$ representing the jerk~\citep{Graefe2022}, balancing simplicity and trajectory smoothness. 
These trajectories are followed by a standard controller~\citep{mellinger2011minimum} acting as $c_i$.
\reviewerall{As our theoretical analysis does not depend on a specific formulation of $\hat{f}_i$, we will continue with an arbitrary $\hat{f}_i$ without giving a specific formula.}

During each round~$k$, a UAV has a predicted reference trajectory $\hat{x}_i(\tau|k)$ and input $\hat{u}_i(\tau|k)$ of the nominal system~\eqref{eq:nominalsystem} computed by a CU, where $\tau$ denotes the time within the prediction horizon. 
With these predictions, the UAV generates the references
\begin{align}
    \hat{x}_i(t) &:= \hat{x}_i(\tau|k), \\
    \hat{u}_i(t) &:= \hat{u}_i(\tau|k),
\end{align}
and follows them using its low-level controller $c_i$ at time $t = \tau + kT$:
\begin{equation}
    u_i(t) = c_i\big[x_i(t),\hat{x}_i(\tau|k),\hat{u}_i(\tau|k)\big].
\end{equation}
When a UAV receives a new trajectory $\hat{u}\uli{i,w}$ from a CU $w$ during the communication phase at round $k$, it updates its reference trajectory at the beginning of round $k+1$. Otherwise, it reuses its old trajectory, either because no CU computed a new trajectory for it or because the UAV missed it due to message loss:
\begin{align} 
	\label{eq:nottriggered} 
	\hat{u}\uli{i}&(\tau|k+1) \\ 
	&= \begin{cases} \hat{u}\uli{i,w}(\tau+T|k) & \parbox[c]{.4\linewidth}{if received new trajectory from a CU $w$,}\\ 
		\hat{u}\uli{i}(\tau+T|k) & \text{otherwise.}\nonumber \end{cases} 
\end{align}

\subsection{MLR-DMPC Algorithm}
\label{sec:mlrdmpc:algorithm}

\definecolor{communicationcolor}{HTML}{fdd48f}
\definecolor{information-trackercolor}{HTML}{d2c0cd}
\definecolor{dmpccolor}{HTML}{bfbfbf}
\definecolor{mlrcolor}{HTML}{8ebae5}

\setlength{\fboxrule}{0pt}
\setlength{\fboxsep}{3pt}

\newcommand{\myindent}{\hspace*{2em}}

\newcommand{\algcolorboxpar}[2]{\vspace*{-\fboxsep}\hspace*{-\fboxsep}\colorbox{#1}{\parbox{\dimexpr\linewidth-2\fboxsep}{#2}}}

\newcommand{\algcolorbox}[2]{%
  {\hskip-\ALG@thistlm\vspace*{-\fboxsep}\hspace*{-\fboxsep}\colorbox{#1}{\parbox{\dimexpr\linewidth-2\fboxsep}{{\hskip\ALG@thistlm\relax {\tiny\strut} #2}}}}%
}

\newcommand{\myif}[1]{\textbf{if} #1 \textbf{then}}
\newcommand{\myelif}[1]{\textbf{elif} #1 \textbf{then}}
\newcommand{\myendif}{\textbf{endif}}
\newcommand{\mywhile}[1]{\textbf{while} #1 \textbf{do}}
\newcommand{\myendwhile}[0]{\textbf{endwhile}}
\newcommand{\varendash}[1][5pt]{%
  \makebox[#1]{\leaders\hbox{--}\hfill\kern0pt}%
}

\begin{algorithm*}[h]
	\caption{MLR-DMPC running on CU $w$.}\label{alg:mlrdmpc}
	\fontsize{9pt}{9pt}\selectfont
	\begin{algorithmic}[1]
		\State\algcolorbox{white}{$k\gets 0$  \Comment{Current round}}
		\State\algcolorbox{white}{$\forall i\in \mathcal{A}$~$\mathcal{D}\uli{iw}(k) \gets \{\}$~~~$\mathcal{D}_w(k)\gets\{\mathcal{D}\uli{iw}(k)|\forall i\in \mathcal{A}\}$ \Comment{Init information-trackers as empty.}}
		\State\algcolorbox{white}{$\forall i\in \mathcal{A}$ setDeprecated($\mathcal{D}\uli{iw}(k)$)}
		
		\State\algcolorbox{white}{state $\gets$ REQUEST\_TRAJECTORY \Comment{Because information-trackers are empty.}}
		
		\State\algcolorbox{white}{UAV\_messages, cu\_messages $\gets \emptyset$}
		
		\State\algcolorbox{white}{\mywhile{true}}
			\State \algcolorbox{information-trackercolor}{\myindent\varendash[0.3\linewidth]\textit{Trajectory-Tracker Update} \varendash[0.3\linewidth]}
			\State  \algcolorbox{information-trackercolor}{\myindent$\mathcal{D}_w(k)\gets$updateInformationTrackers(UAV\_messages, cu\_messages, $\mathcal{D}_w(k-1)$) \label{line:updateinformation-tracker} \Comment{Algorithm~\ref{alg:processmessages}}}  \label{line:computationphase:begin}
			\State \algcolorbox{information-trackercolor}{\myindent\myif{allInformationTrackersUpToDate()} state $\gets$ RUN\_DMPC}
			\State \algcolorbox{information-trackercolor}{\myindent\myelif{state $==$ RUN\_DMPC} state $\gets$ WAIT}\label{lst:line:activatemlr}
			\State \algcolorbox{information-trackercolor}{\myindent\myendif{}}

			\State \algcolorbox{dmpccolor}{\myindent\varendash[0.3\linewidth]\textit{DMPC} \varendash[0.475\linewidth]}
			\State \algcolorbox{dmpccolor}{\myindent\myif{state $==$ RUN\_DMPC}}
				\State \algcolorbox{dmpccolor}{\myindent\myindent prios$\gets$ prioConsensus(cu\_messages) \label{line:prio}\Comment{Event-trigger}}
				\State \algcolorbox{dmpccolor}{\myindent\myindent selected\_UAV $\gets$ selectUAV(prios) \label{line:et}}
				\State \algcolorbox{dmpccolor}{\myindent\myindent tx\_message $\gets$ EmptyMessage}
				\State \algcolorbox{dmpccolor}{\myindent\myindent \myif{$|\mathcal{D}_{\mathrm{selected\_UAV},w}(k)|==1$}} \label{lst:line:onlyoneentrance}
					\State \algcolorbox{dmpccolor}{\myindent\myindent\myindent $\hat{x}_{i,w}(\tau|k),\hat{u}_{i,w}(\tau|k)\gets$ solveOptimization(selected\_UAV,$\mathcal{D}_w(k)$) \label{line:optimization}}
					\State \algcolorbox{dmpccolor}{\myindent\myindent\myindent tx\_message $\gets$ TrajectoryMessage($\hat{x}_{i,w}(\tau|k),\hat{u}_{i,w}(\tau|k)$)} 
				\State \algcolorbox{dmpccolor}{\myindent\myindent\myendif}

			\State \algcolorbox{mlrcolor}{\myindent\varendash[0.3\linewidth]\textit{MLR} \varendash[0.49\linewidth]}
			\State \algcolorbox{mlrcolor}{\myindent\myelif{state $==$ WAIT}   \label{line:mlrbegin}					\Comment{\parbox[t]{.5\linewidth}{Immediately after entering MLR, we do not exactly know which trajectories are unknown.}}} 
				\State \algcolorbox{mlrcolor}{\myindent\myindent tx\_message $\gets$ EmptyMessage}
				\State \algcolorbox{mlrcolor}{\myindent\myindent state $\gets$ REQUEST\_TRAJECTORY}
			\State\algcolorbox{mlrcolor}{\myindent\myelif{state $==$ REQUEST\_TRAJECTORY} \label{line:reqtraj}}
				\State \algcolorbox{mlrcolor}{\myindent\myindent requested\_UAV $\gets$ selectUAVWithDeprecatedInformationTracker($\mathcal{D}_w(k)$)}
				\State \algcolorbox{mlrcolor}{\myindent\myindent tx\_message $\gets$ TrajectoryRequestMessage(requested\_UAV)}
				\State \algcolorbox{mlrcolor}{\myindent\myindent state $\gets$ WAIT\_FOR\_UPDATE}
			\State\algcolorbox{mlrcolor}{\myindent\myelif{state $==$ WAIT\_FOR\_UPDATE}}
				\State \algcolorbox{mlrcolor}{\myindent\myindent tx\_message $\gets \emptyset$
                    \Comment{\parbox[t]{.5\linewidth}{Do not send anything, because requested UAV uses communication resources.}} \label{line:sendnothing}}
				\State \algcolorbox{mlrcolor}{\myindent\myindent state $\gets$ REQUEST\_TRAJECTORY}
			\State \algcolorbox{mlrcolor}{\myindent\myendif{}		 \label{line:mlrend}}	\label{line:computationphase:end}

			\State \algcolorbox{communicationcolor}{\myindent\varendash[0.3\linewidth]\textit{Communication Phase} \varendash[0.3\linewidth]} \label{line:communicationphase:begin}
			\State \algcolorbox{communicationcolor}{\myindent tx\_message $\gets$ \{tx\_message, calcPrio($D_w(k)$)\} \textbf{if} tx\_message $!=$ $\emptyset$ \label{line:calcprio}}
			\State \algcolorbox{communicationcolor}{\myindent UAV\_messages, cu\_messages = wirelessBusRound(tx\_message) \label{line:mixerround}}
			\State  \algcolorbox{communicationcolor}{\myindent$k\gets k+1$} \label{line:communicationphase:end}
		\State \algcolorbox{white}{\myendwhile}
	\end{algorithmic}
\end{algorithm*}

\begin{algorithm*}[t]
	\caption{Algorithm that updates the trajectory information-trackers of the MLR-DMPC.}\label{alg:processmessages}
	\fontsize{9pt}{9pt}\selectfont
	\begin{algorithmic}[1]
		\Procedure{updateInformationTrackers}{UAV\_messages, cu\_messages, $\mathcal{D}_w(k-1)$}
		\State $\forall \mathcal{D}\uli{iw}(k-1)\in \mathcal{D}_w(k-1):$ ~ $\mathcal{\tilde{D}}\uli{iw}(k) \gets \mathcal{D}\uli{iw}(k-1)$ \label{lst:line:metadatacompbegin}
		\For{($i$, message) $\in$ enumerate(UAV\_messages)} \Comment{\parbox[t]{.3\linewidth}{Which trajectories received in the last round}}
		\For{trajectory $\in\mathcal{\tilde{D}}\uli{iw}(k)$}
		\If{trajectory.metadata $==$ message.metadata}
		\State $\mathcal{\tilde{D}}\uli{iw}(k) \gets \{\text{trajectory}\}$ \label{lst:line:update}
		\State setInformationTrackerUpToDate($i$)
		\State \textbf{break}
		\EndIf
		\EndFor
		\EndFor \label{lst:line:metadatacompend}
		\If{$\exists i\in\mathcal{A}$ s.t. $|\mathcal{\tilde{D}}\uli{iw}(k)| > 1$}
		\State setAllInformationTrackersDeprecated() \label{lst:line:deprecatedone}
		\EndIf
		\State \label{lst:line:updateinformationtrackerbegin}
		\State $\mathcal{D}\uli{iw}(k) \gets \mathcal{\tilde{D}}\uli{iw}(k)$
		\If{\textbf{not} length(cu\_messages) == $M$}
		\State setAllInformationTrackersDeprecated() \label{lst:line:deprecatedtwo}
		\EndIf
		\For{(message, $i$) $\in$ enumerate(cu\_messages)}
		\If{type(message) \textbf{is} TrajectoryMessage}
		\State $\mathcal{D}\uli{iw}(k) \gets \mathcal{D}\uli{iw}(k) \cup$\{message\} \label{lst:line:setnewtraj}
		\EndIf
		\EndFor \label{lst:line:updateinformationtrackerend}
		\State \Return $\{\mathcal{D}\uli{iw}(k)|\forall i\in\mathcal{A}\}$
		\EndProcedure
	\end{algorithmic}
\end{algorithm*}


Using MLR-DMPC, the CUs concurrently calculate the trajectories $\hat{u}\uli{i,w}$ for the UAVs. 
Building upon the brief overview presented in Section~\ref{sec:method:mldrmpc}, we now present its details step by step going through Algorithm~\ref{alg:mlrdmpc}.

Each CU alternates between computation (Lines~\ref{line:computationphase:begin}--\ref{line:computationphase:end}) and communication phases (Lines~\ref{line:communicationphase:begin}--\ref{line:communicationphase:end}), with a state machine scheduling operations during computation.

A key component of the computation phase is the information tracker $\mathcal{D}\uli{iw}(k)$, which CU $w$ uses to store trajectory candidates for UAV $i \in \mathcal{A}$ at time $k$, considering one trajectory as currently followed by the UAV.
\reviewerthree{We mark trajectories saved in these trackers with a tilde symbol, e.g., $\tilde{x}$. 
For simplicity and brevity, we abuse the ``element of $\mathcal{D}$'' notation for states, inputs, and positions ambiguously, i.e., we note $\tilde{u}\in \mathcal{D}\uli{iw}(k), \tilde{x}\in \mathcal{D}\uli{iw}(k), \tilde{p} \in \mathcal{D}\uli{iw}(k)$, although $\tilde{u}$, $ \tilde{x}$ and $\tilde{p}$ are of different types and dimensions.}
Due to the delay $T$ between rounds, each trajectory in $\mathcal{D}\uli{iw}(k)$ begins at $k - 1$, e.g., $\tilde{u}_i(\cdot\,|\,k - 1) \in \mathcal{D}\uli{iw}(k)$.

At the start of each computation phase, the CU updates its information trackers (Line~\ref{line:updateinformation-tracker}) using messages from the previous communication phase (Line~\ref{line:mixerround}). 
If it detects critical message loss, it marks the affected trackers as deprecated, indicating they may not contain the UAV's actual trajectory (details in Section~\ref{sec:mlrdmpc:algorithm:informationtracker}).

If all information trackers are up-to-date, the CU executes a DMPC step: selecting a UAV $i$ using the event trigger (Lines~\ref{line:prio} and~\ref{line:et}; Section~\ref{sec:mlrdmpc:algorithm:eventtrigger}) and solving an optimization problem (Line~\ref{line:optimization}) to compute the new reference trajectory $\hat{x}\uli{i,w}(\tau\,|\,k)$ and $\hat{u}\uli{i,w}(\tau\,|\,k)$.
The optimization problem uses the information trackers to formulate anti-collision constraints (details in Section~\ref{sec:mlrdmpc:algorithm:opt}).

If any information tracker is deprecated, the CU initiates MLR (Lines~\ref{line:mlrbegin}--\ref{line:mlrend}) by requesting a missing trajectory from a UAV (Line~\ref{line:reqtraj}). 
In the next round, the UAV transmits its current reference trajectory, using the communication resources typically reserved for the CU because its own are occupied with state information and target position. Accordingly, the CU refrains from transmitting during this round to free communication resources for the UAV (Line~\ref{line:sendnothing}).

At the end of each step, during the communication phase, the CU broadcasts a message containing the MLR-DMPC results and the computed UAV priorities for the next round's event trigger to all other CUs and UAVs (Line~\ref{line:mixerround}), unless it has donated its communication resources to a UAV.

\subsubsection{Information-Tracker Updates}
\label{sec:mlrdmpc:algorithm:informationtracker}

Algorithm~\ref{alg:processmessages} details how each CU updates its information trackers. 
It first extracts the UAVs' current reference trajectories from the received messages. 
Each UAV's message includes unique metadata, specifically, the calculation time and the responsible CU, about its current reference trajectory. 
The CU compares this metadata with that in its information trackers; if they match, it retains the trajectory as the one the UAV is following (Line~\ref{lst:line:update}).

If no new trajectory was planned for a UAV in the previous round, the stored trajectory is the current one. 
However, if a CU did plan a new trajectory, the UAV may be following either the new trajectory (if it received it) or the old one (if it missed the message). 
The CU therefore retains both trajectories in the information tracker (Line~\ref{lst:line:setnewtraj}).

With this information, the CU checks for critical message loss. 
If it received fewer CU messages than the total number of CUs, it might lack the actual trajectory of an unknown UAV, so it marks all information trackers as deprecated (Line~\ref{lst:line:deprecatedtwo}). 
If the CU did not receive a required message from a UAV with necessary metadata to determine which trajectory the UAV followed, it marks that UAV's information tracker as deprecated (Line~\ref{lst:line:deprecatedone}).

\subsubsection{Event-Trigger}
\label{sec:mlrdmpc:algorithm:eventtrigger}

The event-trigger aims to assign $M$ of the $N$ UAVs to the CUs in a distributed manner, avoiding a single point of failure, using priorities similar to~\cite{PredictiveTriggering, Graefe2022}.

At each round $k-1$, each CU $w$ calculates a priority $J\uli{i,w}(k-1)$ for every UAV $i$ (see Algorithm~\ref{alg:mlrdmpc}, Line~\ref{line:calcprio}).
Appendix~\ref{app:et} outlines three triggers with their priority calculation: round-robin (RR) selecting UAVs periodically; distance-based (DB) selecting UAVs farthest from their targets; and a hybrid trigger (HT) combining RR and DB, selecting UAVs based on both their distance to targets and time since last assigned to a CU.

Message loss (e.g., missed target positions) can cause discrepancies among CUs, leading them to compute different priorities for the same UAVs. To address this, CUs employ a consensus algorithm. During communication, they share their computed priorities.
CUs unify them by taking the maximum across all CUs:
\begin{equation}
\label{eq:priosmax}
J\uli{i}(k) = \max_{w} J\uli{i,w}(k-1).
\end{equation}
They sort the UAVs based on these unified priorities and store the top $M$ UAVs in the set $\mathcal{A}_{\mathrm{ET}}(k)$. 
Since all active CUs receive the same priority information, they compute the same set $\mathcal{A}_{\mathrm{ET}}(k)$.

Each CU $w$ selects the UAV at position $((k + w) \bmod M)$ in $\mathcal{A}_{\mathrm{ET}}(k)$\footnote{The modulo operation ensures each priority rank is selected, even when a CU misses multiple messages.}. 
The set of UAVs for which new trajectories are computed is denoted by $\tilde{\mathcal{A}}_{\mathrm{ET}}(k) \subseteq \mathcal{A}_{\mathrm{ET}}(k)$, since some CUs may be in MLR mode and not compute a new trajectory.

\subsubsection{Trajectory Calculation}
\label{sec:mlrdmpc:algorithm:opt}

\reviewerthree{The trajectory computed by CU $w$ for UAV $i$ has a piecewise-constant input with sampling time $T\ul{s}$ and horizon $h\ul{s}$.
The constant input from time $T\ul{s}\kappa$ till $T\ul{s}(\kappa+1)$ for $\kappa\in\{0,...,h\ul{s}-1\}$ is denoted as $u\uli{i,w, \kappa|k}$.
Formally, the trajectory of the system is
\begin{equation}
    \hat{u}\uli{i,w}(\tau+T|k) = \sum_{\kappa=0}^{h\ul{s}-1}\Gamma\ul{T\ul{s}}(\tau-\kappa T\ul{s})u\uli{i,w, \kappa|k},
\end{equation}
where $\Gamma\ul{T\ul{s}}(t) = 1$ if $0 \leq t \leq T\ul{s}$ and $\Gamma\ul{T\ul{s}}(t) = 0$ otherwise~\citep{Graefe2022}. 
Hence, $\Gamma\ul{T\ul{s}}(\tau-\kappa T\ul{s})u\uli{i,w, \kappa|k}=u\uli{i,w, \kappa|k}$, between $T\ul{s}\kappa$ till $T\ul{s}(\kappa+1)$, and it is zero otherwise.
We require $\tfrac{T}{T\ul{s}} = \tfrac{h\ul{s}}{H}$, where $H \in \mathbb{N}$ is the prediction horizon, so the prediction extends to time $kT + HT + T$. 
An offset of $T$ accounts for the latency of one round.
}

\reviewerthree{
The CU aims to determine this input trajectory by minimizing the quadratic distance between the state trajectory and its target, scaled by positive definite matrices $Q\uli{i}$ and $R\uli{i}$. 
This trajectory is constrained by the nominal UAV model $\hat{f}_i$ (Assumption~\ref{as:tracking}). 
The initial state $\tilde{x}\uli{i}(2T|k-1)$ is the sole entry in $\mathcal{D}_{i,w}(k)$ (cf.\ Algorithm~\ref{alg:mlrdmpc}, Line~\ref{lst:line:onlyoneentrance}). 
Both the input and state are confined to $\hat{\mathcal{U}}$ and $\hat{\mathcal{X}}$, respectively, thereby restricting the movement area of the UAV.
Through an equality constraint, the CU ensures that the UAV halts at the end of the computed trajectory, thereby guaranteeing recursive feasibility of the optimization problem~\citep{Graefe2022,Park2022,Chen2022,Chen2023}. Lastly, to ensure collision-free trajectories, the CU imposes time-varying BVC constraints described below.

In summary, the CU solves:

\begin{subequations}
	\label{eq:opt}
	\begin{align}
		\label{eq:objectfunc}
		\min_{\hat{u}\uli{i,w, \cdot|k} }\sum_{\kappa=0}^{h\ul{o}}\big[&||\hat{x}\uli{i,w}(\kappa T\ul{o}+T|k)-\hat{x}\uli{i, \mathrm{target}}||_{Q\uli{i}}^2\nonumber\\
		& + ||\hat{u}\uli{i,w}(\kappa T\ul{o}+T|k)||_{R\uli{i}}^2\big]
	\end{align}
	\begin{align}
		\label{eq:dynconst}
		\text{s.t.~} & \dot{\hat{x}}\uli{i,w}(t|k) = \hat{f}\uli{i}(\hat{x}\uli{i,w}(t|k), \hat{u}\uli{i,w}(t|k))\\
		&\hat{x}\uli{i,w}(T|k)=\tilde{x}\uli{i}(2T|k-1)\nonumber
	\end{align}
	\begin{equation}
		\label{eq:inputconst}
            \hat{u}\uli{i,w, \ell|k} \in \hat{\mathcal{U}} ~\forall\ell\in\left\{0,...,h\ul{s}-1\right\}
	\end{equation}
	\begin{equation}
		\label{eq:stateconst}
		\hat{x}\uli{i,w}(\kappa T\ul{b} + T|k)\in\hat{\mathcal{X}}~
		\forall\kappa\in \{0,...,h\ul{b}\}
	\end{equation}
	\begin{equation}
		\label{eq:feascond}
		0 = \hat{f}\uli{i}(\hat{x}\uli{i,w}(HT+T|k), 0)
	\end{equation}

	\begin{align}
		\forall  j&\in\mathcal{A}\backslash\{i\}:\label{eq:anticollisionconst}\\
		&A\uli{i, j, \mathrm{c}}\begin{bmatrix}
			\hat{p}\uli{i,w}(T|k)\\
			\hat{p}\uli{i,w}(T+T\ul{c}|k)\\
			\vdots\\
			\hat{p}\uli{i,w}(T+h\ul{c}T\ul{c}|k)
		\end{bmatrix} \leq b\uli{i, j, \mathrm{c}}(\mathcal{D}_{j,w}(k))\nonumber
	\end{align}
\end{subequations}
\noindent
where $\hat{x}\uli{i, \mathrm{target}}$ satisfies $p\uli{i, \mathrm{target}} = \hat{g}\uli{\mathrm{p}, i}(\hat{x}\uli{i, \mathrm{target}})$. 
}


Constraint (\ref{eq:anticollisionconst}) prevents collisions between the UAVs.
The matrices $A\uli{i, j, \mathrm{c}}$ and $b\uli{i, j, \mathrm{c}}$ are constructed using time-variant BVC~\citep{Graefe2022,van2017distributed,Chen2022,Chen2023}. 
However, we relax the constraints in case a UAV is not recomputed.
The collision avoidance constraints are computed between the UAV $i$ and every other UAV $j\in\mathcal{A}\backslash\{i\}$.
The CU uses the reference positions $\hat{p}\uli{i}(\cdot|k-1)$ that UAV $i$ is currently following (known because $\mathcal{D}_{i,w}(k)$ contains only one element, cf. Alg.~\ref{alg:mlrdmpc} Line~\ref{lst:line:onlyoneentrance}) and all trajectories $\tilde{p}\uli{j}(\cdot|k-1)\in\mathcal{D}_j(k)$ in the information-tracker for the other UAV $j$.
Thus, the CU constrains the position of UAV $i$ based on all trajectories which it guesses UAV $j$ might fly.
First, we define the difference vector 
\begin{align}
	&n\uli{ij}(hT\ul{c}+2T|k-1)\\
	&= \Theta^{-1}[\tilde{p}\uli{j}(hT\ul{c}+2T|k-1)-\hat{p}\uli{i}(hT\ul{c}+2T|k-1)],\nonumber
\end{align}
which is the normal vector of a plane spanned between the UAVs.
UAV $i$ is constrained to stay on its side of the plane
\begin{align}
	\label{eq:tvb}
	&n\uli{\mathrm{0}, ij}(hT\ul{c}+2T|k-1)^T\Theta^{-1}\\
	&\opindent\times[\tilde{p}\uli{j}(hT\ul{c}+2T|k-1)-\hat{p}\uli{i,w}(hT\ul{c}+T|k)]\nonumber \\
	&\geq \begin{cases}
		\parbox[c]{.55\linewidth}{$\frac{1}{2}(\hat{d}\ul{min}$\\ \hphantom{A}+ $||n\uli{ij}(hT\ul{c}+2T|k-1)||\ul{2})$} & \parbox[c]{.28\linewidth}{if $j\in\mathcal{A}\ul{ET}(k)$}\\
		\hat{d}\ul{min} & \text{else}
	\end{cases} \nonumber
\end{align}
with $n\uli{\mathrm{0}, ij}=n\uli{ij} / ||n\uli{ij}||\ul{2}$ and $\hat{d}\ul{min}$ as the minimum distance between UAVs' reference positions with some safety gap to $d\ul{min}$ (cf. Section~\ref{sec:mdethod:collisionavoidance}).
If UAV $j\in\mathcal{A}\ul{ET}(k)$, we span the plane midway between it and UAV $i$. Else, relaxing the constraints, we move this plane close to $j$, because $j$ remains on its old trajectory.

In this work, we assume a UAV flying space devoid of static obstacles, such as buildings. However, incorporating such obstacles into MLR-DMPC is straightforward using established methods from other DMPCs~\citep{Park2022,Park2023,Chen2023}.

\subsection{Theoretical Analysis}
Our theoretical analysis of \myswarm{} is based on the system model derived under Assumption~\ref{as:tracking}. We first analyze the information trackers $\mathcal{D}$, then demonstrate that the reference trajectories $\hat{p}$ are collision-free, and finally prove the same for the actual positions $p$.

\subsubsection{Content of the Information Trackers}

We present two properties of the information trackers~$\mathcal{D}\uli{iw}(k)$, which follow directly from the construction of Algorithm~\ref{alg:processmessages} and are proven in Appendix~\ref{app:proofs}.
\begin{lemma}
    \label{col:information-tracker}
    If the information-tracker  $\mathcal{D}\ul{iw}(k)$ is not deprecated, then it contains
    the true trajectory $\hat{p}\uli{i}$, the UAV $i$ is following. 
\end{lemma}

\begin{lemma}
    \label{col:information-trackerequal}
    If after executing Algorithm~\ref{alg:processmessages} the two information-trackers $\mathcal{D}\ul{iw}(k)$ and $\mathcal{D}\ul{iv}(k)$ are not deprecated for different $w, v$, then $\mathcal{D}\ul{iw}(k) = \mathcal{D}\ul{iv}(k)$.
\end{lemma}

The CUs hence perform the DMPC step with the same information, which is crucial for guaranteeing collision avoidance.
We thus write $\mathcal{D}\ul{iw}(k) = \mathcal{D}\ul{iv}(k) = \mathcal{D}\ul{i}(k)$ when a CU is not deprecated.
In the following, we treat $\mathcal{D}\ul{i}(k)$ as the global knowledge shared by all CUs executing the DMPC. 
\reviewerone{This simplifies the notation for subsequent derivations.}

When all CUs are in the MLR state, we consider $\mathcal{D}\ul{i}(k)$ as the information tracker of a virtual CU that does not compute new trajectories but listens to all messages without message loss. 
Since no new trajectories are recomputed in this case, the information tracker remains unchanged.

\begin{figure*}[h]
    \centering
    \includegraphics[width=0.9\textwidth]{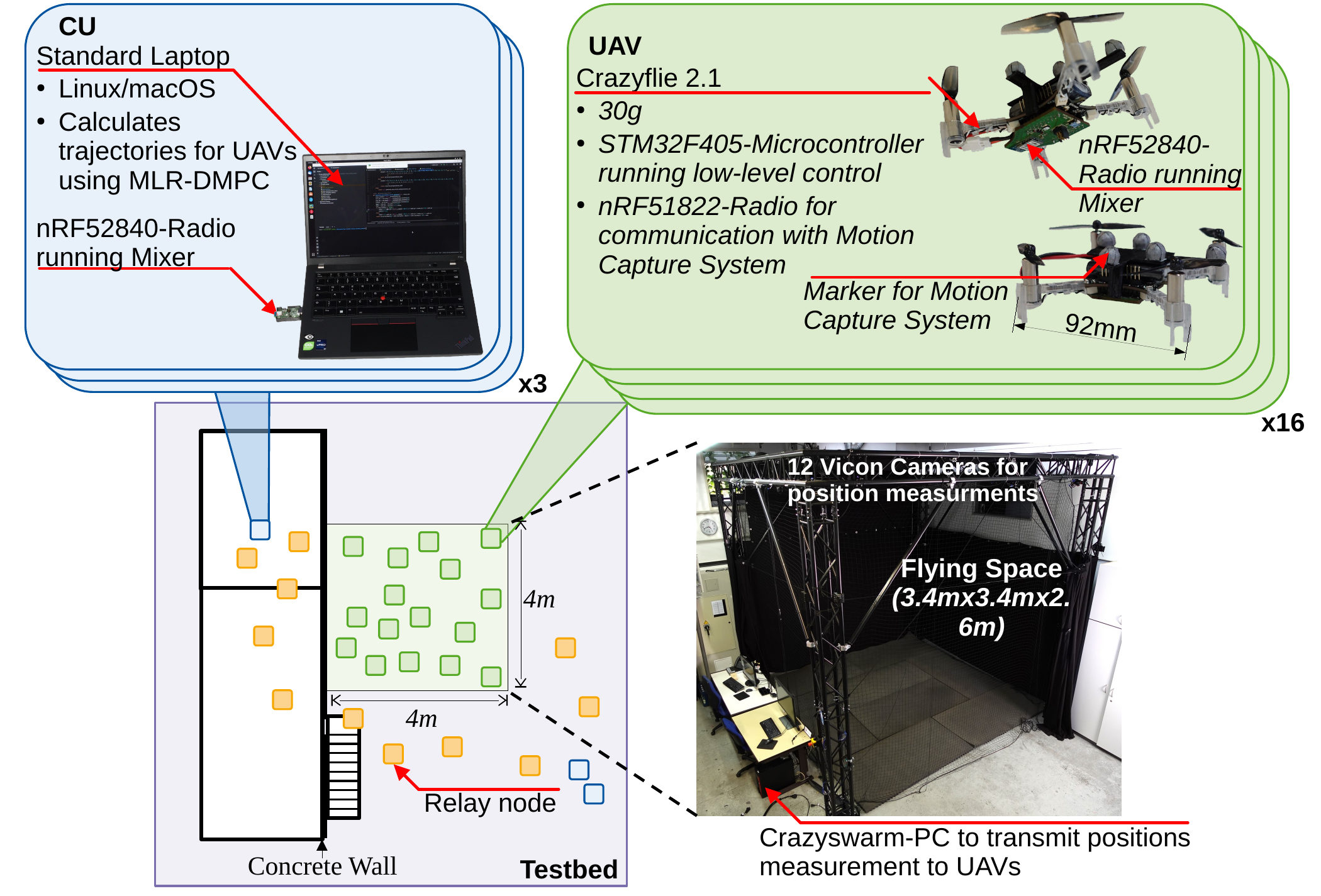}

	\caption{Hardware Implementation of \myswarm{}.
    \capt{Sixteen Crazyflie~2.1 quadcopters operate in a space of $3.4\SI{}{\meter}\times3.4\SI{}{\meter}\times2.6\SI{}{\meter}$. Three laptops serving as CUs compute trajectories using MLR-DMPC. The quadcopters and CUs communicate over a wireless mesh network with at least two hops using the Mixer protocol.
	}}
    \label{fig:hardwareimplementation}
\end{figure*}

\subsubsection{Collision Avoidance of MLR-DMPC}
\label{sec:mdethod:collisionavoidance}
The reference trajectories under constraints (\ref{eq:anticollisionconst}) are collision free as Appendix~\ref{app:proofs} proves.
\begin{corollary}
\label{cor:anitcol}
	If $i,j\in\mathcal{\tilde{A}}\ul{ET}(k)$ and constraint~(\ref{eq:anticollisionconst}) is fulfilled for both $i$ and $j$ on CUs $w$ and $v$, then for all $h\in\{1, 2,\cdots, h_\mathrm{c}\}$ 
	\begin{equation}
		||\Theta^{-1}[\hat{p}\uli{j,v}(hT\ul{c}+T|k)-\hat{p}\uli{i,w}(hT\ul{c}+T|k)]||\ul{2} \geq \hat{d}\ul{min}.
	\end{equation}
\end{corollary}

\begin{lemma}
	\label{lem:anticolone}
	If $i\in\mathcal{\tilde{A}}\ul{ET}(k)$, $j\notin\mathcal{\tilde{A}}\ul{ET}(k)$ and constraint~(\ref{eq:anticollisionconst}) is fulfilled, then for all $\tilde{p}\uli{j}\in\mathcal{D}\ul{j}(k+1)$ and for all $h\in\{1, 2,\cdots, h_\mathrm{c}\}$
	\begin{equation}
		||\Theta^{-1}[\tilde{p}\uli{j}(hT\ul{c}+T|k)-\hat{p}\uli{i,w}(hT\ul{c}+T|k)]||\ul{2} \geq \hat{d}\ul{min}.
	\end{equation}
\end{lemma}

\begin{lemma}
	\label{lem:anticoltwo}
	If for all $i\notin\mathcal{\tilde{A}}\ul{ET}(k)$, $j\notin\mathcal{\tilde{A}}\ul{ET}(k)$, for all $\tilde{p}\uli{j}\in\mathcal{D}\ul{j}(k)$ and $\tilde{p}\uli{i}\in\mathcal{D}\ul{j}(k)$ and for all $h\in\{1, 2,\cdots, h_\mathrm{c}\}$
	\begin{align}
		||\Theta^{-1}[\tilde{p}\uli{j}(hT\ul{c}+T|k-1)&-\tilde{p}\uli{i}(hT\ul{c}+T|k-1)]||\ul{2}\nonumber \\
		&\geq \hat{d}\ul{min},
	\end{align}
	then for all $\tilde{p}\uli{j}\in\mathcal{D}\ul{j}(k+1)$ and $\tilde{p}\uli{i}\in\mathcal{D}\ul{i}(k+1)$ and for all $h\in\{1, 2,\cdots, h_\mathrm{c}\}$
	\begin{equation}
		||\Theta^{-1}[\tilde{p}\uli{j}(hT\ul{c}|k)-\tilde{p}\uli{i}(hT\ul{c}|k)]||\ul{2} \geq \hat{d}\ul{min}.
	\end{equation}
\end{lemma}

We now can prove collision avoidance of all reference trajectories at discrete timepoints.

\begin{theorem}
	\label{th:feas}
	If for all pairwise different $i, j\in\mathcal{A}$, $\forall \tilde{p}\uli{i}\in\mathcal{D}\uli{i}(k)$, $\forall \tilde{p}\uli{j}\in\mathcal{D}\uli{j}(k)$ and $\forall h\in\{0, 1,\cdots, h_\mathrm{c}-1\}$
	\begin{align}
		\label{eq:recfeascond}
		||\Theta^{-1}[\tilde{p}\uli{j}(hT\ul{c}+2T|k-1)&-\tilde{p}\uli{i}(hT\ul{c}+2T|k-1)]||\ul{2}\nonumber\\ 
		&\geq \hat{d}\ul{min},
	\end{align}
	and all $\tilde{x}\uli{i}\in\mathcal{D}\uli{i}(k)$ satisfy constraints~(\ref{eq:dynconst})--(\ref{eq:feascond}), then the optimization problems (\ref{eq:opt}) are feasible for arbitrary message loss.
	Additionally, for all pairwise different $i, j\in\mathcal{A}(k-1)$, $\forall \tilde{p}\uli{i}\in\mathcal{D}\uli{i}(k+1)$, $\forall \tilde{p}\uli{j}\in\mathcal{D}\uli{j}(k+1)$ and for all $h\in\{1, 2,\cdots, h_\mathrm{c}\}$
	\begin{equation}
		\label{eq:colfree}
		||\Theta^{-1}[\tilde{p}\uli{j}(hT\ul{c}+T|k)-\tilde{p}\uli{i}(hT\ul{c}+T|k)]||\ul{2} \geq \hat{d}\ul{min}
	\end{equation}
    and all $\tilde{p}\uli{i}$ also fulfill constraints~(\ref{eq:dynconst})--(\ref{eq:stateconst}).
\end{theorem}
\begin{proof}
	We split the UAVs into two sets. 
	The first set contains all UAVs that were not recomputed, and the second the UAVs that were recomputed. 
	For the first set, we know that the shifted trajectories $\tilde{x}\uli{i}(t+T|k)=\tilde{x}\uli{i}(t+2T|k-1)$ fulfill the constraints~(\ref{eq:dynconst})--(\ref{eq:stateconst}) for all $t<(H-1)T$, as $T$ is a multiple of the sampling times $T\ul{s}$ and $T\ul{b}$~\citep{Graefe2022}. 
	For $t\geq(H-1)T$, it is $\tilde{x}\uli{i}(t+T|k)=\tilde{x}\uli{i}((H-1)T+T|k)=\tilde{x}\uli{i}((H-1)T+2T|k-1)$ and thus also~(\ref{eq:dynconst})--(\ref{eq:stateconst}) and~(\ref{eq:feascond}) are fulfilled. 
	
	For the UAVs that are recomputed, with the same argumentation, we can show that the shifted trajectory $\bar{x}\uli{i}(t+T|k)=\tilde{x}\uli{i}(t+2T|k-1)$ fulfills constraints~(\ref{eq:dynconst})--(\ref{eq:stateconst}).
    Also it holds that for all $j\in\mathcal{A}\backslash\{i\}$ and all $\tilde{p}_j\in\mathcal{D}_j(k)$

    \begin{align}
		n&\uli{\mathrm{0}, ij}(hT\ul{c}+2T|k-1)^T\Theta^{-1}\nonumber\\&\opindent\times[\tilde{p}\uli{j}(hT\ul{c}+2T|k-1)-\bar{p}\uli{i}(hT\ul{c}+T|k)]\nonumber\\
		&=n\uli{\mathrm{0}, ij}(hT\ul{c}+2T|k-1)^T\Theta^{-1}\nonumber\\&\opindent\times[\tilde{p}\uli{j}(hT\ul{c}+2T|k-1)-\tilde{p}\uli{i}(hT\ul{c}+2T|k-1)]\nonumber\\
		&=||\Theta^{-1}[\tilde{p}\uli{j}(hT\ul{c}+2T|k-1)\nonumber\\
		&\opindent-\bar{p}\uli{i}(hT\ul{c}+2T|k-1)]||\ul{2}\nonumber\\
		&\geq\frac{1}{2}(\hat{d}\ul{min}  + ||n\uli{ij}(hT\ul{c}+2T|k-1)||\ul{2}) \nonumber\\
		&\geq \hat{d}\ul{min}.
    \end{align}
    Hence, $\bar{x}$ also fulfills constraint~(\ref{eq:anticollisionconst}) and is a candidate solution optimization problem~(\ref{eq:opt}).
    The optimization problem is thus feasible.
    
    Algorithm~\ref{alg:processmessages} then incorporates the solutions the CUs computed into the information-trackers of every CU after a communication step. 
    First, it deletes some trajectories in the information-trackers (Lines~\ref{lst:line:metadatacompbegin}--\ref{lst:line:metadatacompend}). 
    Because the remaining trajectories were already in the former information-tracker for CUs not in the MLR-state (no deprecated information-tracker), they fulfill the constraints~(\ref{eq:dynconst})--(\ref{eq:feascond}) as argued above. 
    In Lines~\ref{lst:line:updateinformationtrackerbegin}--\ref{lst:line:updateinformationtrackerend}, the algorithm adds new trajectories to the information-tracker, resulting in the information-trackers $\mathcal{D}_i(k+1), \forall~i\in\mathcal{A}$ stored on the CUs with non-deprecated information-trackers.
    These also fulfill constraints~(\ref{eq:dynconst})--(\ref{eq:anticollisionconst}) as they are solutions to the corresponding optimization problem.

    Corollary~\ref{cor:anitcol}, Lemma~\ref{lem:anticolone} and~\ref{lem:anticoltwo} then lead to equation (\ref{eq:colfree}).
\end{proof}

From this, we can derive that the UAVs also fly on collision-free trajectories.

\begin{theorem}
	\label{th:collisionfreeness}
	If the reference trajectories of all UAVs fulfill~(\ref{eq:dynconst})--(\ref{eq:feascond}) at $k=0$ and Equation (\ref{eq:recfeascond}), then for all $k$, arbitrary message loss, all pairwise different $i, j\in\mathcal{A}$ and and for all $h\in\{0, 1,\cdots, h_\mathrm{c}-1\}$
	\begin{equation}
		||\Theta^{-1}[\hat{p}\uli{j}(hT\ul{c}+kT)-\hat{p}\uli{i}(hT\ul{c}+kT)]||_2 \geq \hat{d}_\mathrm{min}.
	\end{equation}
\end{theorem}
\begin{proof}
First, we apply induction to Theorem~\ref{th:feas}, which is possible as the non-deprecated information-trackers are all equal to $\mathcal{D}_{i}$ due to Lemma~\ref{col:information-trackerequal}, and $T$ is a multiple of $T_\mathrm{c}$~\citep{Graefe2022}.
Thus, for pairwise different $i, j\in\mathcal{A}$, the information-tracker $\mathcal{D}_{i}$ contains trajectories, which are collision-free to all other trajectories in the information-tracker $\mathcal{D}_{j}$. 
Because of Lemma~\ref{col:information-tracker}, each UAV follows one trajectory in the information-tracker, and its reference positions are collision-free.
\end{proof}

Based on Assumption~\ref{as:tracking}, we now derive guarantees for the actual UAV positions at continuous times $t$.

\begin{theorem}
\label{th:colfreerealworld}
    Let
        \begin{align}
        \label{eq:contopt}
            \Delta d_\mathrm{min,cont}&=\hat{d}_\mathrm{min}\\&-\min_{\substack{\hat{x}_{0, i},\hat{x}_{0, j}\in\hat{\mathcal{X}}\\ \tau\in[0,T_\mathrm{c}]\\\hat{u}_i(t), \hat{u}_j(t)\in\hat{\mathcal{U}}}}||\Theta^{-1}(\hat{p}_i(\tau)-\hat{p}_j(\tau))||_2\nonumber\\
        \text{s.t.~~} \dot{x}_i&=\hat{f}_i(\hat{x}_i(t), \hat{u}_i(t)), \hat{x}_i(0)=\hat{x}_{0,i}\nonumber\\
        \dot{x}_j&=\hat{f}_i(\hat{x}_j(t), \hat{u}_j(t)), \hat{x}_j(0)=\hat{x}_{0,j}\nonumber\\
        \hat{p}_i(t) &= g_i(\hat{x}_i(t))\nonumber\\
        \hat{p}_j(t) &= g_i(\hat{x}_j(t))\nonumber\\
        ||\Theta^{-1}(\hat{p}_i(0)&-\hat{p}_j(0))||_2 \geq \hat {d}_\mathrm{min}\nonumber\\
        ||\Theta^{-1}(\hat{p}_i(T_\mathrm{c})&-\hat{p}_j(T_\mathrm{c}))||_2 \geq \hat {d}_\mathrm{min}.\nonumber\\
        \end{align}
    If
    \begin{equation}
        \hat{d}_\mathrm{min} \geq d_\mathrm{min} + \Delta d_\mathrm{min} + \hat{d}_\mathrm{min,cont},
    \end{equation}
    then Equation (\ref{eq:truecollisionavoidance}) holds, i.e., the UAVs are collision free.
\end{theorem}

\begin{proof}
    Follows directly from Assumption (\ref{eq:truecollisionavoidance}), construction of $\Delta d_\mathrm{min,cont}$ and triangle inequality.
\end{proof}

For collision avoidance guarantees, $\Delta d_\mathrm{min}$ and $\Delta d_\mathrm{min,cont}$ must be known. 
Solving optimization problem~(\ref{eq:contopt}) directly yields $\Delta d_\mathrm{min,cont}$, while $\Delta d_\mathrm{min}$ must be determined through experiments or engineering intuition.

\begin{remark}
Under strong external disturbances like gusts, Assumption~\ref{as:tracking} may not hold. 
If a UAV's state deviates significantly from the reference, we reinitialize its current state and assign a high priority $J\uli{i}$ to ensure replanning~\citep{Luis2020}. 
However, under these conditions, the guarantees of MLR-DMPC do not hold, as such disturbances may lead the UAV onto a collision course.
\end{remark}

\reviewerone{
\begin{remark}
Theorem~\ref{th:colfreerealworld} requires collision-free initial trajectories. 
In practice, UAVs typically start their flight by hovering while maintaining a safe distance from one another. 
Therefore, the initial trajectories can often be initialized as this constant hovering state.
\end{remark}
}

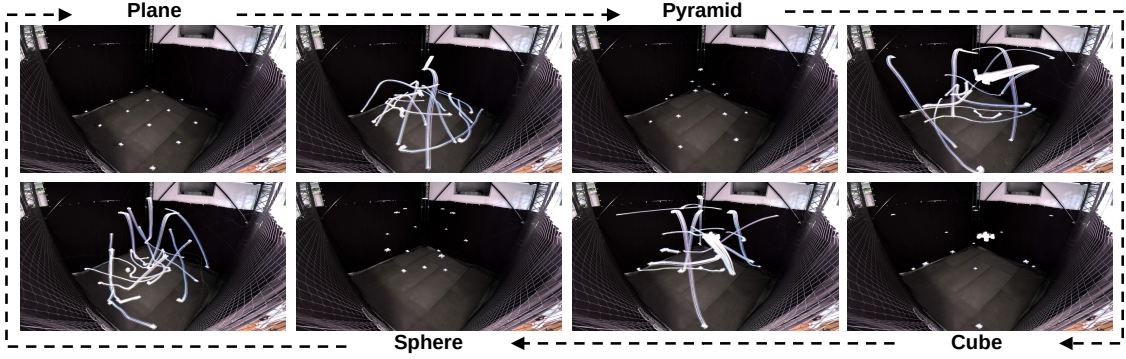
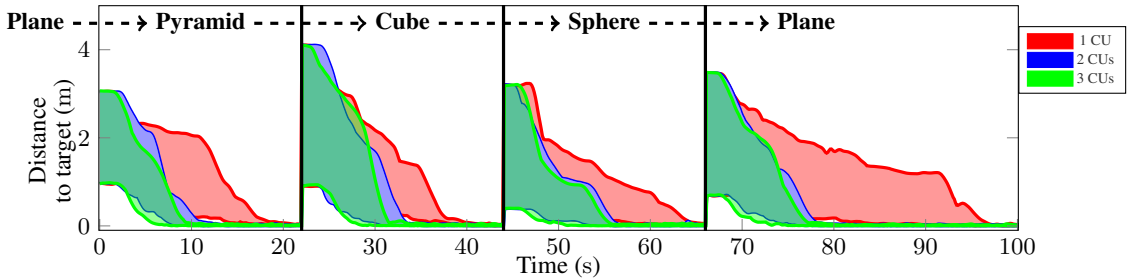
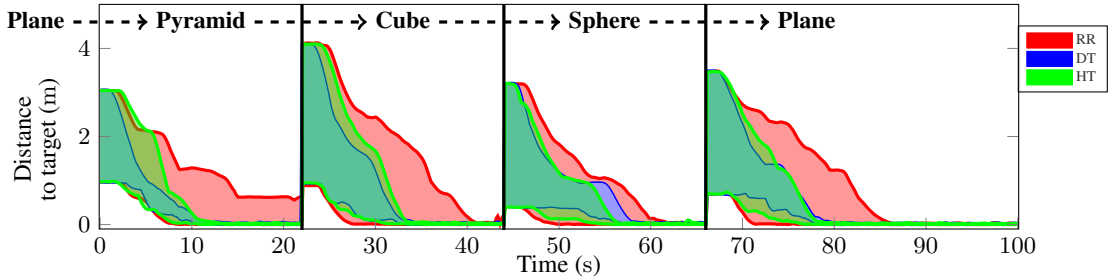
\begin{figure*}[t]
	\centering
	\fontsize{9.3pt}{9.3pt}\selectfont
	\newcommand{\radiusplot}{0.1}
	\newcommand{\plotlinewidth}{1.0pt}
	\newcommand{\myfigwidth}{0.4\textwidth}
	\newcommand{\myfigheight}{0.3\textwidth}
	\newcommand{\marksize}{1.0pt}
	\newcommand{\marksizedrone}{3.0pt}
	\newcommand{\boxshift}{0.05cm}
	\newcommand{\boxsize}{0.05cm}
	\newcommand{\constraintlinewidth}{1pt}
	\newcommand{\lw}{1.2pt}
	\newcommand{\marksizequadcopter}{2.0pt}
	\newcommand{\colormlrdmpc}{green!50!black}

	\newcommand{\colorone}{red}
	\newcommand{\colortwo}{blue}
	\newcommand{\colorthree}{green}

	\newcommand{\newtasklinewidth}{1.2pt}
	\newcommand{\labelheight}{4.6}

    \begin{subfigure}{0.95\textwidth}
        \includegraphics[width=0.99\linewidth]{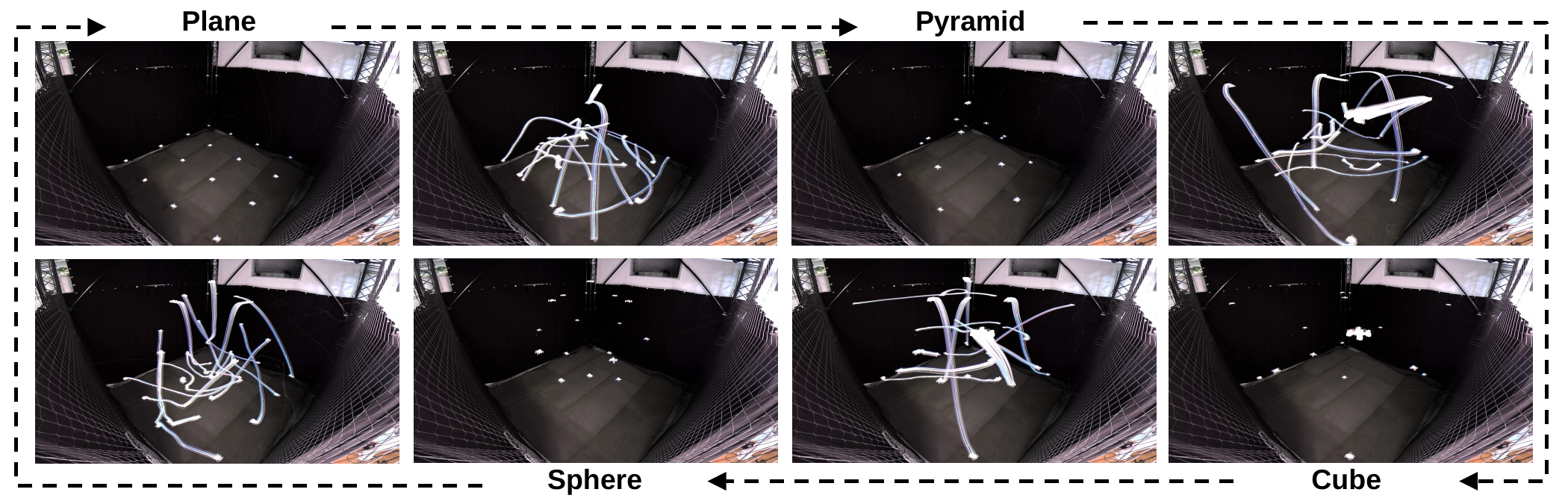}
        \caption{The swarm's maneuvers in the experiments.}
        \label{fig:hardware:distrcomp:maneouver}
    \end{subfigure}

	\begin{subfigure}{0.95\textwidth}
		\centering
		\begin{tikzpicture}[spy using outlines={circle, magnification=3, size=2cm, connect spies, every spy on node/.append style={thick}}]
			\begin{axis}[
                    axis on top,
                    fill between/on layer={main},
                    xmax=100.1, xmin=-0.1,
                    ymax=5, ymin=-0.1,
                    xlabel={Time (\SI{}{\second})},
                    xlabel style={yshift=3mm},
                    xtick pos=bottom, 
                    ylabel near ticks,
                    ylabel style={align=center, xshift=-3mm, yshift=-2mm}, ylabel={Distance\\ to target (\SI{}{\meter})},
                    legend style={at={(axis cs: 100.1,3)},anchor=south west ,draw=black,fill=white,align=left, fill opacity=0.8, nodes={scale=0.6, transform shape}}, 
                    height=\myfigheight, width=0.9\textwidth, box plot width=0.15cm,
				]
					
                    \addplot[mark=none, color=\colorone, line width=\lw, name path=lowertemp, forget plot] table [x=t, y=dmin, col sep=comma] {plot_data/HardwareExperimentFigures_1CUs.csv};
                    
                    \addplot[mark=none, color=\colorone, line width=\lw, name path=uppertemp, forget plot] table [x=t, y=dmax, col sep=comma] {plot_data/HardwareExperimentFigures_1CUs.csv};
                    
                    \addplot[color=\colorone, opacity=\fillopacity, legend image post style={opacity=1.0}] fill between[of=uppertemp and lowertemp];
                    \addlegendentry{1 CU};



                \addplot[mark=none, color=\colortwo, line width=\lw, name path=lowertempp, forget plot] table [x=t, y=dmin, col sep=comma] {plot_data/HardwareExperimentFigures_2CUs.csv};
                
                \addplot[mark=none, color=\colortwo, line width=\lw, name path=uppertempp, forget plot] table [x=t, y=dmax, col sep=comma] {plot_data/HardwareExperimentFigures_2CUs.csv};

                \addplot[color=white, legend image post style={opacity=1.0}, forget plot] fill between[of=uppertempp and lowertempp];
                
                \addplot[color=\colortwo, opacity=\fillopacity, legend image post style={opacity=1.0}] fill between[of=uppertempp and lowertempp];
                \addlegendentry{2 CUs};



                \addplot[mark=none, color=\colorthree, line width=\lw, name path=lowertemppp, forget plot] table [x=t, y=dmin, col sep=comma] {plot_data/HardwareExperimentFigures_3CU.csv};
                
                \addplot[mark=none, color=\colorthree, line width=\lw, name path=uppertemppp, forget plot] table [x=t, y=dmax, col sep=comma] {plot_data/HardwareExperimentFigures_3CUs.csv};

                \addplot[color=\colorthree, opacity=\fillopacity, legend image post style={opacity=1.0}] fill between[of=uppertemppp and lowertemppp];
                \addlegendentry{3 CUs};

				
                \draw[solid, line width=\newtasklinewidth] (axis cs:22,-2) -- (axis cs:22,8);
                \draw[solid, line width=\newtasklinewidth] (axis cs:44,-2) -- (axis cs:44,8);
                \draw[solid, line width=\newtasklinewidth] (axis cs:66,-2) -- (axis cs:66,8);

                \coordinate (planecoordone) at (axis cs: -7, \labelheight);
                \coordinate (pyramidcoord) at (axis cs: 11, \labelheight);
                \coordinate (cubecoord) at (axis cs: 33, \labelheight);
                \coordinate (spherecoord) at (axis cs: 55, \labelheight);
                \coordinate (planecoord) at (axis cs: 77, \labelheight);

			\end{axis}

                \node[anchor=center] (planeone) at (planecoordone) {\small \textbf{Plane}\strut};
                \node[anchor=center] (pyramid) at (pyramidcoord) {\small \textbf{Pyramid}\strut};
                
                \node[anchor=center] (cube) at (cubecoord) {\small \textbf{Cube}\strut};
                
                \node[anchor=center] (sphere) at (spherecoord) {\small \textbf{Sphere}\strut};
                
                \node[anchor=center] (plane) at (planecoord) {\small \textbf{Plane}\strut};
                
                \draw[dashed, line width=\newtasklinewidth, ->] (planeone) -- (pyramid);
                \draw[dashed, line width=\newtasklinewidth, ->] (pyramid) -- (cube);
                \draw[dashed, line width=\newtasklinewidth, ->] (cube) -- (sphere);
                \draw[dashed, line width=\newtasklinewidth, ->] (sphere) -- (plane);

            \end{tikzpicture}
            \caption{Performance of the swarm with respect to the number of CUs.}
            \label{fig:hardware:distrcomp:numcu}
        \end{subfigure}

        \begin{subfigure}{0.95\textwidth}
            \centering
            \begin{tikzpicture}[spy using outlines={circle, magnification=3, size=2cm, connect spies, every spy on node/.append style={thick}}]
            \begin{axis}[
                    axis on top,
                    fill between/on layer={main},
                    xmax=100.1, xmin=-0.1,
                    ymax=5.0, ymin=-0.1,
                    xlabel={Time (\SI{}{\second})}, 
                    xlabel style={yshift=3mm},
                    xtick pos=bottom,
                    ylabel near ticks,
                    ylabel style={align=center, xshift=-3mm, yshift=0mm}, ylabel={Distance\\ to target (\SI{}{\meter})},
                    legend style={at={(axis cs: 100,3)},anchor=south west ,draw=black,fill=white,align=left, fill opacity=0.8, nodes={scale=0.6, transform shape}}, 
                    height=\myfigheight, width=0.9\textwidth, box plot width=0.15cm,
				]
					
                    \addplot[mark=none, color=\colorone, line width=\lw, name path=lowertemp, forget plot] table [x=t, y=dmin, col sep=comma] {plot_data/HardwareExperimentFigures_RR.csv};
                    
                    \addplot[mark=none, color=\colorone, line width=\lw, name path=uppertemp, forget plot] table [x=t, y=dmax, col sep=comma] {plot_data/HardwareExperimentFigures_RR.csv};
                    
                    \addplot[color=\colorone, opacity=\fillopacity, legend image post style={opacity=1.0}] fill between[of=uppertemp and lowertemp];
                    \addlegendentry{RR};
                    


				\addplot[mark=none, color=\colortwo, line width=\lw, name path=lowertempp, forget plot] table [x=t, y=dmin, col sep=comma] {plot_data/HardwareExperimentFigures_DT.csv};

				\addplot[mark=none, color=\colortwo, line width=\lw, name path=uppertempp, forget plot] table [x=t, y=dmax, col sep=comma] {plot_data/HardwareExperimentFigures_DT.csv};

				\addplot[color=white, legend image post style={opacity=1.0}, forget plot] fill between[of=uppertempp and lowertempp];

				\addplot[color=\colortwo, opacity=\fillopacity, legend image post style={opacity=1.0}] fill between[of=uppertempp and lowertempp];
				\addlegendentry{DT};



				\addplot[mark=none, color=\colorthree, line width=\lw, name path=lowertempp, forget plot] table [x=t, y=dmin, col sep=comma] {plot_data/HardwareExperimentFigures_2CUs.csv};

				\addplot[mark=none, color=\colorthree, line width=\lw, name path=uppertempp, forget plot] table [x=t, y=dmax, col sep=comma] {plot_data/HardwareExperimentFigures_2CUs.csv};

				\addplot[color=\colorthree, opacity=\fillopacity, legend image post style={opacity=1.0}] fill between[of=uppertempp and lowertempp];
				\addlegendentry{HT};


                \draw[solid, line width=\newtasklinewidth] (axis cs:22,-2) -- (axis cs:22,8);
                \draw[solid, line width=\newtasklinewidth] (axis cs:44,-2) -- (axis cs:44,8);
                \draw[solid, line width=\newtasklinewidth] (axis cs:66,-2) -- (axis cs:66,8);

                \coordinate (planecoordone) at (axis cs: -7, \labelheight);
                \coordinate (pyramidcoord) at (axis cs: 11, \labelheight);
                \coordinate (cubecoord) at (axis cs: 33, \labelheight);
                \coordinate (spherecoord) at (axis cs: 55, \labelheight);
                \coordinate (planecoord) at (axis cs: 77, \labelheight);

			\end{axis}

			\node[anchor=center] (planeone) at (planecoordone) {\small \textbf{Plane}\strut};
			\node[anchor=center] (pyramid) at (pyramidcoord) {\small \textbf{Pyramid}\strut};

			\node[anchor=center] (cube) at (cubecoord) {\small \textbf{Cube}\strut};

			\node[anchor=center] (sphere) at (spherecoord) {\small \textbf{Sphere}\strut};

			\node[anchor=center] (plane) at (planecoord) {\small \textbf{Plane}\strut};

			\draw[dashed, line width=\newtasklinewidth, ->] (planeone) -- (pyramid);
			\draw[dashed, line width=\newtasklinewidth, ->] (pyramid) -- (cube);
			\draw[dashed, line width=\newtasklinewidth, ->] (cube) -- (sphere);
			\draw[dashed, line width=\newtasklinewidth, ->] (sphere) -- (plane);

		\end{tikzpicture}
        \caption{Performance of the swarm with respect to the event-trigger.}
        \label{fig:hardware:distrcomp:et}
	\end{subfigure}

    \caption{Hardware experiment results on distributed computation.
    \capt{
    \textbf{(a)} Visualization of flown formations.
    \textbf{(b)} Swarm performance improves with more CUs, illustrating the trade-off between resource efficiency and performance.
    \textbf{(c)} Comparison of event triggers: RR periodically selects all UAVs; DT selects based on proximity to targets; HT combines both. RR performs worst overall, while DT and HT each excel in different scenarios. A video of the experiments is available at \url{http://tiny.cc/DMPCSwarmComputation}.
    }}

	\label{fig:hardware:distrcomp}
\end{figure*}

\begin{figure*}[t]
    \centering
    \includegraphics[width=0.99\textwidth]{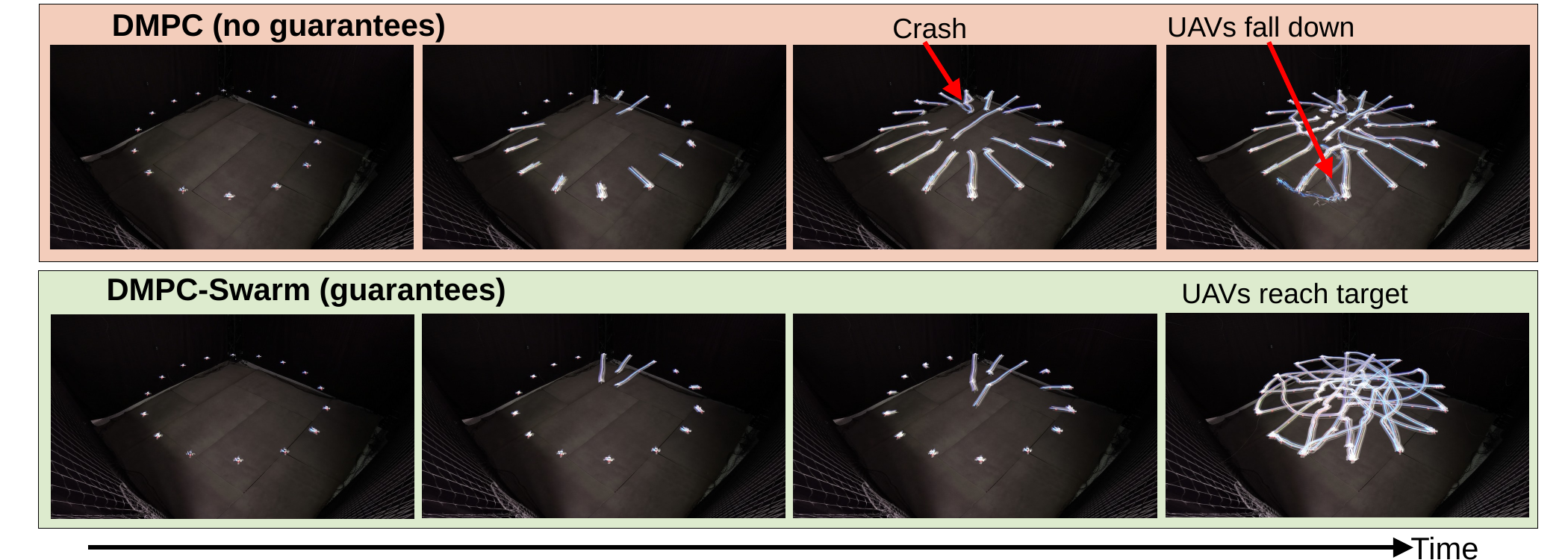}
    \caption{Comparison of MLR-DMPC with DMPC ignoring message loss. \capt{Sixteen UAVs fly to the opposite side of a circle. We jam the wireless channel for the first two seconds. Using existing methods~\citep{Graefe2022} that ignore message loss, the UAVs crash. With MLR-DMPC, the UAVs remain safe. A video of the maneuver is available at \url{http://tiny.cc/DMPCSwarmMessageLoss}.}}
    \label{fig:hardware:dmpcvsmlrdmpc}
\end{figure*}

\section{Experimental Results}
\label{sec:experiments}



We demonstrate \myswarm's ability to safely control real UAV swarms using DMPC. Specifically, we present:

\begin{enumerate}
    \item The overall performance of \myswarm{} as the first implementation of DMPC with distributed computation and wireless communication (Section~\ref{sec:experiments:mainresult}).
    \item An evaluation of \myswarm{}'s distributed event-triggered computation architecture (Section~\ref{sec:experiments:distrcomp}).
    \item \myswarm{}'s robustness under message loss (Section~\ref{sec:experiments:messageloss}).
\end{enumerate}

\subsection{Experimental Setup}
\label{sec:method:hardware}

Using the described architecture as a base, we achieve a distributed DMPC implementation (see Figure~\ref{fig:hardwareimplementation}). 
It includes 16 Crazyflie~2.1 quadcopters and three laptops as CUs, which could correspond to end-user devices in an application scenario.
All devices are equipped with nRF52840 \SI{2}{\mega\bit\per\second} BLE transceivers running Mixer.

The swarm operates within a $3.4 \times 3.4 \times 2.6$~\SI{}{\meter} obstacle-free space. 
A Vicon motion capture system with Crazyswarm~\citep{preiss2017crazyswarm} provides position measurements, which in real application could be substituted by distributed solutions such as Ultra-Wideband or GPS localization.

\reviewerone{We chose $\hat{d}_\mathrm{min}=\SI{0.25}{\meter}$ through experimentation.
We began with $\hat{d}_\mathrm{min}=\SI{0.35}{\meter}$ from~\cite{Luis2019}. 
Through further testing, we found that reducing the distance to $\hat{d}_\mathrm{min}=\SI{0.25}{\meter}$ allowed the swarm to operate reliably. 
Distances smaller than this proved unfeasible not due to physical collisions between UAVs, but because rotor-generated turbulence reduced lift, leading to drones crashing into the ground unexpectedly.}

By positioning the CUs approximately 15 meters apart and separating them with a concrete wall to prevent direct communication, the network has at least two hops. 
The MLR-DMPC frequency is \SI{5}{\hertz} ($T_\mathrm{calc} = \SI{105}{\milli\second}$, $T_\mathrm{com}=\SI{95}{\milli\second}$), while the Crazyflies' low-level control runs at \SI{500}{\hertz}. 

\subsection{Performance of \myswarm{}}
\label{sec:experiments:mainresult}
Our video (\url{http://tiny.cc/DMPCSwarm}) showcases key experimental outcomes, including various swarm maneuvers. 
It demonstrates that \myswarm{} successfully implements DMPC-based swarm control in hardware, executing the entire DMPC algorithm in a distributed manner across the quadcopters and CUs.

Despite challenges such as changing wireless properties due to robot movement, communication among up to 27 devices in a mesh network, and interference from external Wi-Fi and Bluetooth devices, the communication network remains functional and all devices stay synchronized.  
The event-trigger mechanism enables dynamic swarm control while conserving resources, even when there are more drones than CUs. 
Importantly, no collisions occur during any maneuvers, despite network latency and message loss.




\subsection{Distributed Event-Triggered Computation}
\label{sec:experiments:distrcomp}

We evaluated our approach through two experiments: (1) examining the influence of the number of CUs (i.e., available computational power), and (2) assessing the effect of the event trigger on swarm performance. 
Both experiments used the same setup (Figure~\ref{fig:hardware:distrcomp:maneouver}): 16 quadcopters underwent multiple formation changes, starting from a planar configuration and sequentially forming a pyramid, cube, sphere, and finally returning to the original plane. 
Results are presented in Figure~\ref{fig:hardware:distrcomp} and in a video (\url{http://tiny.cc/DMPCSwarmComputation}).


\fakepar{Number of CUs}
We control the swarm using one, two, and three CUs. As shown in Figure~\ref{fig:hardware:distrcomp:numcu}, increasing the number of CUs accelerates the formation changes. 
The most significant improvement occurs when increasing from one to two CUs, while adding a third CU yields diminishing returns. 
Thus, adding computational power enhances swarm performance.

Using fewer CUs conserves resources but reduces performance; the extent of this trade-off depends on the number of CUs.

\fakepar{Event-Trigger}
We can influence swarm performance by choosing how event-trigger priorities are calculated (cf.\ Section~\ref{sec:mlrdmpc:algorithm:eventtrigger})~\citep{Graefe2022}. We compare RR, DB and HT.

Figure~\ref{fig:hardware:distrcomp:et} presents the results. 
In three of the four formations, the swarm using RR is the slowest to reach targets because RR often assigns resources to agents already at their targets. 
DB and HT generally exhibit similar performance across formations. 
With HT, some quadcopters reach targets earlier than others; with DB, all quadcopters reach targets at approximately the same time, as DB exclusively assigns resources to those farthest from their targets.

The event-trigger enables dynamic control of the swarm, even with many more drones than CUs. 
Selecting the appropriate event-trigger is crucial for achieving optimal performance.

\subsection{Robustness under Message Loss}
\label{sec:experiments:messageloss}

To highlight the importance of \myswarm{}'s collision avoidance under message loss, we compare MLR-DMPC with a DMPC that does not account for message loss~\citep{Graefe2022}.

In this experiment, sixteen quadcopters controlled by three CUs exchange their positions along a circle's circumference (see Figure~\ref{fig:hardware:dmpcvsmlrdmpc} and \url{http://tiny.cc/DMPCSwarmMessageLoss}). 
To make our experiment reproducable, we simulate a jammed communication channel, introducing a two-second period during which the CUs experience message loss---they can transmit but cannot receive data.

We conduct the experiment twice: first with CUs running MLR-DMPC, then with standard DMPC. 
With standard DMPC, two quadcopters collide, causing the swarm to fail its task. 
In contrast, under MLR-DMPC control, all quadcopters remain safe and successfully exchange positions without collision, demonstrating the critical role of our MLR module.

\reviewerone{We note that \myswarm{} would have been safe for arbitrarily long message-loss periods. In this case, the UAVs would have come safely to a halt after the length of the MPC's prediction horizon.}

\section{Conclusion}

We introduced \myswarm{}, the first hardware implementation of DMPC-based swarm control with distributed computation and wireless mesh communication. 
\myswarm{} employs the communication protocol Mixer, showing that its synchronized many-to-all communication structure, robust to rapid device movement, is ideal for realizing DMPC. 
\myswarm{} uses ground-based, event-triggered distributed computations on CUs, enabling flexible scaling with the number of UAVs and efficient resource usage, even with limited bandwidth and hardware resources. 
Finally, we developed a novel DMPC algorithm that provably prevents collisions even under message loss. 
Our hardware experiments demonstrate that this combination enables DMPC on nano UAV swarms for the first time.

Our findings on the suitability of synchronous transmission protocols for UAV swarm communication extend beyond DMPC. 
Future research could explore the benefits of this communication architecture for other UAV swarm control methods.

\reviewerthree{Although our distributed control approach is efficient for small and medium-sized swarms, it does not readily scale to huge swarms of hundreds to thousands of agents.
First, the communication rounds would take too long due to the many-to-all structure.
Second, solving the optimization problems would also take a long time as MLR-DMPC includes all UAVs in anti-collision constraints.
In such large-scale systems, a many-to-all communication approach is not necessary.
For example, \cite{Chen2022} showed that for DMPC, UAVs only need to consider other UAVs in their neighborhood.
Although directly integrating this idea into MLR-DMPC is feasible, there is currently no efficient means of achieving local many-to-many communication via synchronous transmissions, which is an interesting challenge for future work.}

Another interesting future challenge is to eliminate Assumption~\ref{as:tracking}. 
For strong external disturbances, like sudden gusts, this Assumption~\ref{as:tracking} is most likely not valid or too conservative (i.e., $\Delta d_\mathrm{min}$ would be very large).
One possible way to eliminate this assumption would be to include recent event-triggered robust MPC~\citep{grafe2025event}.

\section*{Acknowledgments}
This work was supported by the German Research Foundation (DFG) within the priority program 1914 (grant TR 1433/2) and within the Emmy Noether project NextIoT (ZI 1635/2-1), and by the LOEWE initiative (Hesse, Germany) within the emergenCITY center (LOEWE/1/12/519/03/05.001(0016)/72). 

The authors gratefully acknowledge the computing time provided to them at the NHR Center NHR4CES at RWTH Aachen University (p0022034). This is funded by the Federal Ministry of Education and Research, and the state governments participating on the basis of the resolutions of the GWK for national high performance computing at universities (www.nhr-verein.de/unsere-partner).

We thank Sebastian Giedyk for his help with the quadcopters and the testbed, and Shengsi Xu for his help with the software development for the experiments, and Fabian Mager, Henrik Hose, Alexander von Rohr and Pierre-François Massiani for helpful discussions.

\bibliography{references}

\begin{appendices}

\section{Event Triggers}
\label{app:et}



To minimize communication overhead in the event-trigger mechanism, we quantize priorities to 8-bit unsigned integers instead of 32-bit floats. 
Due to the one-round delay, a recalculated UAV's priority may not reflect its updated state. 
To address this, if a CU has just recalculated UAV~$i$, it sets its priority to zero. 
We adjust the maximum operation in Equation~(\ref{eq:priosmax}) to return zero if any element is zero; otherwise, it returns the maximum. The CU calculates priorities using the following equations.

\subsection{Round-Robin Event Trigger (RR)}
The RR calculates priorities as:
\begin{equation}
    J_{iw}(k) = k - k_{i, \text{calc}}(k),
\end{equation}
where $k_{i, \text{calc}}(k)$ is the last round in which UAV $i$'s trajectory was calculated. If the CU's database contains multiple trajectories, it uses the first one.

\subsection{Distance-Based Event Trigger (DT)}
The DT calculates priorities as:
\begin{equation}
    J_{iw}(k) = \left\| p_{i, \mathrm{target}} - p_i\big(T\,|\,(k - 1)T\big) \right\|.
\end{equation}

\subsection{Hybrid Event Trigger (HT)}
The HT combines the previous methods:
\begin{align}
    J_{iw}(k) &= \left\| p_{i, \mathrm{target}} - p_i\big(T\,|\,(k - 1)T\big) \right\|\nonumber\\ 
    &\opindent\times\left[ k - k_{i, \text{calc}}(k) \right].
\end{align}

\subsection{Deadlock-Aware Triggering}

Since recalculating a deadlocked UAV's trajectory will not change it, we introduce a deadlock-aware triggering mechanism. 
The CU uses \cite[Theorem~1, Equation~(13)]{Chen2022} to detect deadlocks. 
If a UAV is in deadlock, the CU sets its priority to one.
\section{Deadlock Avoidance}
\label{app:deadlock}

The main idea behind these constraints is to incorporate rotation into the swarm constraints~\citep{Chen2023}. 
To achieve this, we modify the left side of constraint~(\ref{eq:tvb}):

\begin{align}
    \label{eq:tvbsoftconstraints}
    &n\uli{\mathrm{0}, ij}(hT\ul{c}+2T|k-1)^T\Theta^{-1} \\
    &\opindent\times[\tilde{p}\uli{j}(hT\ul{c}+2T|k-1)-\hat{p}\uli{i,w}(hT\ul{c}+T|k)] \nonumber \\
    &\geq \begin{cases}
        \parbox[c]{.55\linewidth}{$\frac{1}{2}(\hat{d}\ul{min}$\\ \hphantom{A}+ $||n\uli{ij}(hT\ul{c}+2T|k-1)||\ul{2})$\\ \hphantom{A}+ $\epsilon$} & \text{if~} j\in\mathcal{A}\ul{ET}(k)\\
        \hat{d}\ul{min} + \epsilon & \text{else}
    \end{cases},\nonumber
\end{align}
where $\epsilon\in \mathbb{R}_{\geq 0}$ is an additional optimization variable.
\reviewerone{$\epsilon \geq 0$ hereby ensures that the constraints in the limit of $\epsilon=0$ are the same as the constraints (\ref{eq:tvb}), ensuring that the guarantees still hold~\cite{Chen2023}}

Following~\cite{Chen2023}, we increase the weight of $\epsilon$ in the objective function when UAV $j$ is to the right of UAV $i$, pushing each UAV to its right and causing approaching UAVs to rotate around each other. 
Unlike~\cite{Chen2023}, we keep the soft constraint hyperparameters constant, as they perform well without adaptation.

\subsection{High-Level Path Planner}

We combine the soft constraints with a high-level planner similar to~\cite{Park2022}. 
In normal operation, the desired target state $x\uli{i, \mathrm{target}}$ in optimization problem~(\ref{eq:opt}) equals the UAV's actual target. 
When a deadlock occurs---detected when all velocities fall below a threshold—a CU activates the high-level planner, and $x\uli{i, \mathrm{target}}$ is set to an intermediate target calculated by the planner.

Each CU runs its portion of the high-level planner for its assigned subset of UAVs in parallel with the communication round, avoiding additional computation time. 
The intermediate target positions are communicated in the subsequent communication round.

The high-level planner classifies UAVs into those continuing towards their targets and those that should make room for others. 
For each assigned UAV~$i$, the CU determines if it should make room for another UAV~$j$ based on the following conditions:
\begin{itemize}
    \item UAV~$i$ is within a certain distance of UAV~$j$,
    \item UAV~$j$ is not closer to its target than UAV~$i$,
    \item UAV~$i$ is moving towards UAV~$j$, or UAV~$i$ lies between UAV~$j$ and its target.
\end{itemize}
The CU notes all UAVs for which UAV~$i$ should make room and adjusts UAV~$i$'s path for the closest such UAV. 
Following~\cite{Park2022}, UAV~$i$ sets its new target to its current position plus a vector pointing away from UAV~$j$, adding random noise to resolve symmetric deadlocks.

UAVs follow these intermediate targets until the deadlock resolves or the conditions necessitating making room no longer apply. 
The high-level planner then stops, and all UAVs resume flying to their actual target positions.

\section{Additional Proofs}
\label{app:proofs}

\begin{proof}[Proof of Lemma 1]
    If UAV $i$ was not recalculated in the previous round, then $|\mathcal{D}_{iw}| = 1$. If the CU received the trajectory metadata, it directly knows the trajectory UAV $i$ is following. If the metadata was not received due to message loss, since the database is not deprecated (it would have been marked otherwise), the CU still knows the trajectory from prior rounds.

    If UAV $i$ was recalculated in the previous round, then at Line~\ref{lst:line:deprecatedtwo} in Algorithm~\ref{alg:processmessages}, the database contains the trajectory UAV $i$ was following two rounds ago. After processing, it includes this trajectory and the new one calculated and transmitted in the last round. Depending on whether UAV $i$ received the new trajectory, it is following one of these two trajectories. Thus, the CU knows the possible trajectories UAV $i$ may be following.
\end{proof}

\begin{proof}[Proof of Lemma 2]
    For UAVs not recalculated in the last round, the intermediate databases $\tilde{\mathcal{D}}_{iv}(k)$ and $\tilde{\mathcal{D}}_{iw}(k)$ contain their actual trajectories (Lines~\ref{lst:line:metadatacompbegin}--\ref{lst:line:metadatacompend}, Lemma~\ref{col:information-tracker}). For UAVs recalculated in the last round, the databases contain their trajectories from the last or second-to-last round, depending on recalculations due to the event trigger. After updating $\tilde{\mathcal{D}}$, the newly calculated trajectories are added (Lines~\ref{lst:line:updateinformationtrackerbegin}--\ref{lst:line:updateinformationtrackerend}). The databases are not deprecated only if all new trajectories are received, ensuring $\mathcal{D}_{iw}(k) = \mathcal{D}_{iv}(k)$.
\end{proof}

\begin{proof}[Proof of Corrolary 1]
    As both UAVs are selected for computation, we know $|\mathcal{D}\uli{j}| = 1$.
    The corollary then follows from Lemma 1 in \citep{Graefe2022}.
\end{proof}
    
\begin{proof}[Proof of Lemma 3] 
    We know that because UAV $j$ was not recalculated $\tilde{p}\uli{j}(\cdot|k)=p\uli{j}(\cdot+T|k-1)\in\mathcal{D}\ul{j}(k-1)$ (Equation (\ref{eq:nottriggered})). 
    It is
    \begin{align}
        ||&\Theta^{-1}[\tilde{p}\uli{j}(hT\ul{c}+T|k)-\hat{p}\uli{i,w}(hT\ul{c}+T|k)]||\ul{2}\nonumber\\ 
        &=||n\uli{\mathrm{0}, ij}(hT\ul{c}+2T|k-1)||\ul{2}||\Theta^{-1}\nonumber\\
        &\opindent\times[\tilde{p}\uli{j}(hT\ul{c}+T|k)-\hat{p}\uli{i,w}(hT\ul{c}+T|k)]||\ul{2}\nonumber \\
        &\geq n\uli{\mathrm{0}, ij}(hT\ul{c}+2T|k-1)^T\Theta^{-1}\nonumber\\
        &\opindent\times[p\uli{j}(hT\ul{c}+T|k)- \hat{p}\uli{i,w}(hT\ul{c}+T|k)]\nonumber\\
        &= n\uli{\mathrm{0}, ij}(hT\ul{c}+2T|k-1)^T\Theta^{-1}\nonumber\\
        &\opindent\times[p\uli{j}(hT\ul{c}+2T|k-1)- \hat{p}\uli{i,w}(hT\ul{c}+T|k)]\nonumber\\
        &\geq \hat{d}\ul{min}
    \end{align}
\end{proof}

\begin{proof}[Proof of Lemma 4] 
Since no new entries are added to the databases, there exists a trajectory $\tilde{p}\uli{i}(\cdot+T|k-1)\in\mathcal{D}_{i}(k)$ with $\tilde{p}\uli{i}(\cdot|k-1)=\tilde{p}\uli{i}(\cdot+T|k)$ (same for $j$), and thus collision freeness follows directly.
\end{proof}

\end{appendices}

\end{document}